\documentclass[preprint,aps,showpacs,nofootinbib,preprintnumbers,amsmath,amssymb,superscriptaddress]{revtex4-2}
\usepackage{amsmath,amssymb,amsthm,mathrsfs}
\usepackage{epsfig}
\usepackage{graphicx}
\usepackage[usenames,dvipsnames]{color}
\usepackage{subfigure}
\usepackage{slashed}
\usepackage{comment}
\usepackage[normalem]{ulem}
\usepackage{tikz}
\usepackage[compat=1.1.0]{tikz-feynman}
\usepackage[colorlinks,citecolor=blue]{hyperref}
\usepackage{color}

\usepackage{multirow}
\tikzfeynmanset{compat=1.1.0}

\usepackage{tikzsymbols}

\usepackage{xcolor}
\usepackage{soul}

\begin{document}
\title{Gravitational Wave Generation and Detection \\ in Gravitational Quantum Field Theory}
\author{Yuan-Kun~Gao\footnote{gaoyuankun17@mails.ucas.ac.cn}}
\affiliation{University of Chinese Academy of Sciences (UCAS), Beijing 100049, China}
\author{Da~Huang\footnote{dahuang@bao.ac.cn}}
\affiliation{National Astronomical Observatories, Chinese Academy of Sciences, Beijing, 100012, China}
\affiliation{School of Fundamental Physics and Mathematical Sciences, Hangzhou Institute for Advanced Study, UCAS, Hangzhou 310024, China}
\affiliation{International Centre for Theoretical Physics Asia-Pacific, Beijing/Hangzhou, China}
\author{Yue-Liang~Wu\footnote{ylwu@ucas.ac.cn,ylwu@itp.ac.cn}}
\affiliation{University of Chinese Academy of Sciences (UCAS), Beijing 100049, China}
\affiliation{School of Fundamental Physics and Mathematical Sciences, Hangzhou Institute for Advanced Study, UCAS, Hangzhou 310024, China}
\affiliation{International Centre for Theoretical Physics Asia-Pacific, Beijing/Hangzhou, China}
\affiliation{Institute of Theoretical Physics, Chinese Academy of Sciences, Beijing 100190, China}

\date{\today}
\begin{abstract}
\noindent We investigate the production and detection of gravitational waves (GWs) within the framework of Gravitational Quantum Field Theory (GQFT). In this theory, GWs exhibit five propagating modes: one scalar, two vector, and two tensor modes. Unlike General Relativity, the gravitational field equations in GQFT involve both symmetric and antisymmetric tensors, governed by their respective energy-momentum tensors, both of which can act as sources for GW radiation. By solving the linearized gravitational equations, we derive general analytic expressions for the different GW degrees of freedom. Our analysis reveals that the symmetric energy-momentum tensor generates scalar and tensor GWs through the trace and traceless parts of the quadrupole moment, respectively. In contrast, the antisymmetric stress tensor induces scalar and vector GWs with enhanced coupling strengths. We examine two illustrative examples: a black hole binary with a slightly elliptical orbit, which produces scalar GWs, and a neutron star binary where one component has a net spin aligned with its velocity, leading to vector GW emission. Finally, we study the detectability of these GW polarizations by analyzing their signatures in GW detectors. Our findings indicate that current observatories can detect both scalar and tensor modes, while a newly designed detector would be required to probe vector GWs.
\end{abstract}

\maketitle
\newpage

\section{Introduction}\label{s1}

Currently, the standard theory describing gravity is Einstein’s General Relativity (GR), which has successfully passed nearly all experimental tests from Solar System scales to cosmological observations (for recent reviews, see {\it e.g.} Refs.~\cite{Will:2014kxa, Will:2018bme} and references therein). However, due to the unresolved challenges in quantizing GR, it is often regarded as a low-energy effective field theory of gravity (for a recent review, see {\it e.g.} Ref.~\cite{Donoghue:2022eay} and references therein). The fundamental nature of gravity remains one of the most profound open questions in modern physics. Numerous attempts have been made to uncover a more fundamental theory underlying GR, including superstring theory~\cite{Green:1987sp,Green:1987mn,Polchinski:1998rr}, loop quantum gravity~\cite{Rovelli:2008zza,Ashtekar:2021kfp}, and other approaches~\cite{Horava:2009uw}. On the other hand, the direct observation of the gravitational waves (GWs) emitted from the merger of binary black holes by the advanced LIGO Collaboration~\cite{LIGOScientific:2016aoc} has marked the beginning of the era to employ GWs as a probe to test various aspects of gravity~\cite{LIGOScientific:2016lio}.     

Gravitational Quantum Field Theory (GQFT)~\cite{Wu:2015wwa,Wu:2022mzr,Wu:2025abi,Wu:2024mul} is a novel framework for gravity based on the gauge symmetry principles. Distinct from traditional gauge theories of gravity~\cite{Hehl:1976kj,Ivanenko:1983fts,Hehl:1994ue} (with comprehensive comparisons provided in ref.~\cite{Wu:2024mul}), GQFT seeks to reconcile GR with Quantum Field Theory (QFT) by postulating that the laws of nature are governed by the intrinsic properties of matter’s fundamental constituents, such as leptons and quarks in the Standard Model. This approach necessitates a clear distinction between intrinsic symmetries (encoded in the quantum numbers of fields) and external symmetries (describing their motion in flat Minkowski spacetime). Consequently, GQFT treats the global Lorentz symmetry SO(1,3) of Minkowski spacetime and the intrinsic spin symmetry SP(1,3) of fermionic fields as a unified joint symmetry SO(1,3)$\Join$SP(1,3), rather than as an associated symmetry in conventional QFT. To preserve this joint symmetry, GQFT introduces a spin-associated bi-covariant vector field $\hat{\chi}_{a}^{\; \, \mu}(x)$, which transforms homogeneously under both spin gauge symmetry SP(1,3) and global Lorentz symmetry SO(1,3). Its dual field $\chi_{\mu}^{\; \, a}(x)$ emerges as the fundamental gravitational field, replacing the metric field $\chi_{\mu\nu}=\eta_{ab}\chi_{\mu}^{\; a}\chi_{\nu}^{\; b}$ of GR. This gravigauge field $\chi_{\mu}^{\; \,a}$ acts as a Goldstone-type boson, identified as the massless graviton, and serves as a gauge-type field defined in Minkowski spacetime and valued in a spin-associated gravigauge spacetime. Remarkably, GQFT provides a framework to unify all fundamental interactions into a hyperunified field theory (see Ref.~\cite{Wu:2022aet} for a comprehensive analysis and the references therein). 

So far, GQFT has been explored in many aspects, including astrophysics~\cite{Gao:2024ihk,Chen:2024bho}, cosmology~\cite{Wu:2022aet,Wang:2023hsb}, and particle physics~\cite{Huang:2022jpl}. Recently, a general theory of the standard model\cite{Wu:2025abi} was developed to integrate the two established standard models in particle physics and cosmology, offering comprehensive insights into the mysteries of the universe's dark sector. In Ref.~\cite{Gao:2024juf}, we examined the linearized gravitational dynamics of GQFT in the absence of matter sources, which reveals that, in addition to the standard tensor polarizations, the theory permits an extra scalar mode and two vector modes as gravitational wave (GW) degrees of freedom (dofs). Given the potential significance of these additional polarizations, we aim to investigate GW production and detection in GQFT.

Similar to GR, GWs in GQFT are sourced by the energy-momentum (EM) tensor, but here the EM tensor is constructed from fundamental quantum fields. A key distinction is that the EM tensor in GQFT contains both symmetric and antisymmetric components~\cite{Wu:2022mzr}. Clarifying the physical meaning of these components and their role in GW generation is a central objective of this work. Furthermore, we assess the prospects for detecting these GW modes with current and future detectors~\cite{Xu:2025yrn}. To this end, we analyze the influence of different polarizations on GW experiments by solving the associated geodesic equations~\cite{Carroll:2004st}.

The paper is structured as follows. In Sec.~\ref{SecRev}, we briefly review the linearized gravitational equations in the presence of both symmetric and antisymmetric EM tensors, establishing notation conventions for later use.
Sec.~\ref{SecRadG} investigates the generation of gravitational waves (GWs) of different polarizations by analytically solving the linearized gravitational field equations. We also explore the physical interpretation of the corresponding sources, by providing two concrete examples of scalar and vector GW radiation. In Sec.~\ref{SecGWDet}, we examine the feasibility of detecting the predicted GW polarizations with present experimental setups. Finally, we summarize our main results and discuss potential limitations in Sec.~\ref{SecConc}. The paper includes three appendices: Appendix~\ref{uDef} explicitly derives the source for vector GW production and interprets its physical meaning in the nonrelativistic limit. Appendix~\ref{SDef} calculates the scalar GW source arising from the antisymmetric stress tensor, focusing on the fermions in the presence of background gauge fields, including the spin gauge field. Appendix~\ref{Proofs} provides detailed proofs of several key identities used in deriving the general solutions to the scalar and vector gravitational equations.

\section{Brief Review of Linearized Gravitational Field Equations in the GQFT }\label{SecRev}

In the GQFT, the fundamental field governing gravidynamics is the gravigauge field $\chi^a_{\mu}$~\cite{Wu:2022mzr,Wu:2025abi}. Upon fixing the local spin gauge symmetry SP(1,3) within this framework, the gravitational interaction is completely characterized by the symmetric component of $\chi^a_\mu$. This symmetric component exhibits identical degrees of freedom to those of the metric field $\chi_{\mu\nu} \equiv \eta_{ab}\chi^a_\mu \chi^b_\nu$.

We decompose the gravigauge field into a Minkowski background and a gravitational perturbation:
\begin{eqnarray}
\chi_\mu^a = \eta^a_\mu + h^a_\mu/2,
\end{eqnarray}
which leads to the following expansion of the metric field $\chi_{\mu\nu}$:
\begin{eqnarray}
	\chi_{\mu\nu} \approx \eta_{\mu\nu} + h_{\mu\nu}\,,
\end{eqnarray}
where $h_{\mu\nu} \equiv \eta_{ab} (h^a_\mu \delta^b_\nu + h^b_\nu \delta^a_\mu)/2$ represents the linearized perturbation field at leading order. The components of $h_{\mu\nu}$ can be further decomposed according to their spin properties, as shown in~\cite{Bertschinger:1993xt,Mukhanov:1990me,Nojiri:2020pqr}: 
\begin{itemize}
	\item Spin-2 tensor modes: $\hat{h}_{ij}$
	\begin{eqnarray}
		\hat{h}_{ij}\,, \quad \hat{h}_{it} = \hat{h}_{tt} =0 \,,\quad \hat{h}^i_i =0\,, \quad \partial^i \hat{h}_{ij} = 0\,,
	\end{eqnarray}
	\item Spin-1 vector modes: $S_i$ and $F_i$
	\begin{eqnarray}
		h_{tt} = 0\,,\quad h_{it} = S_i\,,\quad h_{ij} = 2\partial_{(i}F_{j)}\,,\quad \partial^i S_i = \partial^iF_i =0\,,
	\end{eqnarray}
	\item Spin-0 scalar modes: $\phi$, $B$, $\psi$ and $E$        
	\begin{eqnarray}
		h_{tt} = -2\phi\,, \quad h_{it} = -\partial_i B\,,\quad h_{ij} = \delta_{ij}(-2\psi) + 2 \partial_i \partial_j E\,.  
	\end{eqnarray}
\end{itemize}
At low energies, we can view $\chi_{\mu\nu}$ as an effective metric tensor, in which its perturbation fields can be summarized with the following compact line element
\begin{eqnarray}\label{Metric}
	ds^2 = - (1+2\phi) dt^2 + 2 (S_i -\partial_i B) dx^i dt + [\hat{h}_{ij} + (1-2\psi)\delta_{ij} + 2 \partial_{(i} F_{j)} + 2 \partial_i \partial_j E ]dx^i dx^j\,.
\end{eqnarray}
Here and in what follows, we shall use the most plus sign convention for the metric tensor $\eta_{\mu\nu} = \mathrm{diag}(-1.1,1,1)$, which is different from earlier works in the GQFT. 

It was pointed out in Ref.~\cite{Gao:2024juf} that the linearized gravitational equations of the GQFT suffers a scalar-type gauge symmetry with the transformations as $\delta h_{\mu\nu} = \partial_\mu\partial_\nu \zeta$, under which various scalar components are modified as
\begin{eqnarray}
	\phi \to \phi - \partial_t^2 \zeta/2\,, \quad B\to B-\partial_t \zeta\,, \quad E\to E+ \zeta/2\,.
\end{eqnarray}
It turns out that we can obtain the following two gauge-invariant variables~\cite{Gao:2024juf}
\begin{eqnarray}
	\Phi = \phi-\partial_t B /2\,,\quad A = B + 2 \partial_t E\,.
\end{eqnarray}

Note that the complete linearized gravitational dynamics in the GQFT is determined by two sets of equations
\begin{eqnarray}\label{GWt}
	\widetilde{G}_{(\mu \nu)} = -8\pi \widetilde{G}_N T_{(\mu\nu)}\,,\quad \widetilde{G}_{[\mu\nu]} = -8\pi \widetilde{G}_N T_{[\mu\nu]}\,.
\end{eqnarray}
which correspond to the symmetric and antisymmetric parts of the complete gravitational equations, respectively. The parameter $\widetilde{G}_N$ is the fundamental gravitational coupling in the GQFT, which can be related to the conventional Newtons constant $G_N$ as~\cite{Gao:2024juf}
\begin{eqnarray}\label{DefGN}
	{G}_N \equiv \frac{1-\gamma_W}{(1-\gamma_W/2)(1+\gamma_W)} \widetilde{G}_N\,.
\end{eqnarray} 
which can be yielded by comparing the Newtonian potential generated by a static mass source in the QGFT with observations. 
Also, we have explored in Ref.~\cite{Gao:2024juf} the source-free gravitational equations with $T_{(\mu\nu)} = T_{[\mu\nu]} = 0$ and found five massless propagating dofs: two spin-2 modes $\hat{h}_{ij}$, two vector modes $S_i$, and one scalar $\psi$. In this paper, we would like to explore the generation and detection of these five GW polarizations. 

Before investigating the GW production in the GQFT, we need to firstly examine the EM tensors shown in Eq.~\eqref{GWt}. Notice that the total EM tensor $T_{\mu\nu}$ is the summation over the symmetric part $T_{(\mu\nu)}$ and the antisymmetric one $T_{[\mu\nu]}$. Thus, in linearized gravitational field approximation, the EM conservation equation in the GQFT is given by
\begin{eqnarray}\label{EMCt}
	\partial^\mu T_{\mu\nu} = \partial^\mu (T_{(\mu\nu)}+T_{[\mu\nu]}) =0\,.
\end{eqnarray}
Similar to the treatment of GW fields $h_{\mu\nu}$, we shall also decompose the source fields into the following spin basis~\cite{Bertschinger:1993xt,Carroll:2004st}:
\begin{eqnarray}\label{DefEMW}
	T_{(00)} = \rho\,, \quad T_{(0i)} = v_i -\partial_i b\,, \quad T_{(ij)} = \sigma_{ij} + p\delta_{ij} + 2 \partial_{(i}\theta_{j)} + 2\partial_i \partial_j q \,,
\end{eqnarray}
and
\begin{eqnarray}\label{DefEMA}
	T_{[0i]} = u_i -\partial_i s \,,\quad T_{[ij]} = \epsilon_{ijk} (\omega^k -\partial^k \lambda)\,. 
\end{eqnarray}
where $v_i$, $u_i$, $\omega_i$ and $\theta_i$ are transverse vectors while $\sigma_{ij}$
is a transverse-traceless tensor. 

Hence, the linearized gravitational equations in Eq.~\eqref{GWt} can be rewritten in terms of these components fields as follows
\begin{eqnarray}
	\widetilde{G}_{(00)} &=& -\partial^i \partial_i (2\psi+ \gamma_W\Phi) = -8\pi \widetilde{G}_N \rho\,, \label{EqG00} \\
	\widetilde{G}_{(0i)} &=& \frac{1}{2}\left[\partial^k \partial_k (S_i -\partial_t F_i)-4\partial_t \partial_i \psi \right] \nonumber\\
	&& +\frac{\gamma_W}{2} \left[\frac{1}{2}\square S_i + \frac{1}{2} \partial_k \partial^k (S_i -\partial_t F_i) - \partial_t \partial_i (\Phi-\psi+ \partial_t A/2) - \frac{1}{2} \partial_t \square A \right] \nonumber\\
	&=& -8\pi\widetilde{G}_N (v_i - \partial_i b) \label{EqG0i} \\
	\widetilde{G}_{(ij)} &=&  \frac{1}{2}\left\{ (1+\gamma_W) \square \hat{h}_{ij} + \gamma_W\square\partial_{(i}F_{j)} + (\gamma_W+2) \partial_t \partial_{(i}[S-\partial_t F]_{j)}  \right. \nonumber\\
	&& + 2 \partial_i \partial_j [(\gamma_W-1)\psi + \Phi -(\gamma_W+1)\partial_t A/2] \nonumber\\
	&& +  \left. \delta_{ij} [ (4-2\gamma_W)\square \psi - 2 \partial^k \partial_k(\Phi+\psi-\partial_t A/2)] \right\} \nonumber\\
	&=& -8\pi\widetilde{G}_N (\sigma_{ij} + p\delta_{ij} + 2\partial_{(i}\theta_{j)} + 2\partial_i \partial_j q)\,. \label{EqGij}
\end{eqnarray} 
and
\begin{eqnarray}
	\widetilde{G}_{[0i]} &=& -\frac{\gamma_W}{4} \left[\square S_i -\partial_k \partial^k (S_i -\partial_t F_i) - 2\partial_t \partial_i (\Phi + \psi -\partial_t A/2) + \partial_i \square A \right] \nonumber\\
	&=& -8\pi \widetilde{G}_N (u_i- \partial_i s)\,,\label{EqH0i} \\
	\widetilde{G}_{[ij]} &=& -\frac{\gamma_W}{2} \left(\square \partial_{[i}F_{j]} - \partial_t \partial_{[i}[S-\partial_t F]_{j]} \right) = -8\pi\widetilde{G}_N\epsilon_{ijk} (\omega^k - \partial^k \lambda)\,. \label{EqHij} 
\end{eqnarray}  

Similarly, we can also express the EM conservation conditions in Eq.~\eqref{EMCt} in terms of component source fields
\begin{eqnarray}
	&\nu = 0&\,: \quad -\partial_t \rho - \partial_i \partial^i (b-s) =0\,,\label{EqC0} \\
	&\nu=i&\,: \quad -\partial_t (v_i-\partial_i b) + \partial_i p + \partial^j \partial_j \theta_i + 2 \partial_i \partial_j \partial^j q - \partial_t (u_i-\partial_i s) - \epsilon_{ijk} \partial^j \omega^k = 0\,,\label{EqCi}
\end{eqnarray}
Note that Eq.~\eqref{EqCi} can be further simplified. By acting a spatial derivative $\partial^i$ on it, we can extract the following equations involving only scalars
\begin{eqnarray}
	\partial^i \partial_i (\partial_t b + p + 2 \partial^j \partial_j q + \partial_t s ) = 0\,,
\end{eqnarray}
which can be reduced to
\begin{eqnarray}\label{EqCs}
	\partial_t b + p + 2 \partial^j \partial_j q + \partial_t s =0\,.\label{EqCiS}
\end{eqnarray}
By further taking this scalar relation back into Eq.~\eqref{EqCi}, we can obtain the following equation
\begin{eqnarray}\label{EqCiV}
	\partial_t (v_i+u_i) + \epsilon_{ijk}\partial^j \omega^k -\partial^j \partial_j \theta_i =0\,.
\end{eqnarray}

It is worth noting that, for a GW source localized in a compact region ${\cal V}$, we can obtain from Eqs.~\eqref{EqC0} and \eqref{EqCi} the following two conserved quantities 
\begin{eqnarray}\label{DefMK}
	M = \int_{\cal V} d^3y \rho \,,\quad K_i = \int_{\cal V} d^3y (v_i + u_i)\,,
\end{eqnarray}
which can be proved by directly integrating Eqs.~\eqref{EqC0} and \eqref{EqCiV} over $\mathcal{V}$ as
\begin{eqnarray}\label{ProofMK}
	\frac{dM}{dt} =  \int_{\cal V} d^3 y  \partial_t\rho = 0 \,,\quad \frac{dK_i}{dt} = \int_{\cal V} d^3y \partial_t(u_i+v_i) = 0\,.
\end{eqnarray}
However, one can show that the constant $K_i$ vanishes as
\begin{eqnarray}\label{ProofK0}
	K_i = \int_{\cal V} d^3y (v_i + u_i) = \int_{\cal V} d^3y (\partial^j y_i) (v_j +u_j) = -\int_{\cal V} d^3 y y^i \partial^j (v_j+u_j) = 0\,. 
\end{eqnarray}
Note that the quantity $K_i$ can be understood as the total momentum of the whole matter system, and its zero value is the consequence of the transversality of vectors $u_i$ and $v_i$. We also assumed that all of the surface terms vanish identically since all matters are confined in a compact region. 
On the other hand, one can construct another conserved vectorial quantity with $(v_i + u_i)$:
\begin{eqnarray}\label{DefL}
	L^i = \int_{\cal V} d^3 y \epsilon^{ijk} y_j (v_k + u_k)
\end{eqnarray}
whose conservation can be  easily proved as
\begin{eqnarray}\label{ConstL}
	&& \frac{dL^i}{dt} =  \int_{\cal V} d^3 y \epsilon^{ijk} y_j \partial_t(v_k+ u_k) = \int_{\cal V} d^3y \epsilon^{ijk} y_j \left(\partial^m \partial_m \theta_k - \partial_{kmn} \partial^m \omega^n\right) \nonumber\\
	&=& -\int_{\cal V} d^3y \left( 2 \omega^i + \epsilon^{ijk} \partial_j \theta_k\right) =0\,.
\end{eqnarray}
Actually, $L^i$ corresponds to the angular momentum of the system. In the next section, the conserved quantities $M$ and $L^i$ are shown to play important roles in the solution to the linearized GQFT equations.  


\section{Gravitational Wave Radiations in the GQFT}\label{SecRadG}

In this subsection, we focus on the GW generation within the framework of GQFT. To achieve this, we solve the complete set of linearized gravitational equations given in Eqs.~\eqref{EqG00}–\eqref{EqHij}, accounting for both the symmetric and antisymmetric components of the EM tensor.

\subsection{Scalar Sector}\label{SecRadS}

Our analysis begins with the scalar sector, which comprises the fields $\Phi$, $\psi$, and $A$. By extracting the scalar components from the general linearized gravitational equations, we derive the dynamical equations governing these fields.

First, we examine Eq.~\eqref{EqH0i} for the antisymmetric components $\widetilde{G}_{[0i]}$,
\begin{eqnarray}
	\partial^i \widetilde{G}_{[0i]} = -({\gamma_W}/{4}) \partial^i \partial_i  \left[ -2 \partial_t (\Phi + \psi-\partial_t A/2) +  \square A \right] = 8\pi \widetilde{G}_N \partial^i \partial_i s\,,
\end{eqnarray} 
which leads to
\begin{eqnarray}\label{EqH0iS}
	(\gamma_W/2) \left[ \partial_t (\Phi+\psi-\partial_t A/2) - \square A/2 \right] = 8\pi \widetilde{G}_N s\,.
\end{eqnarray}
Similar argument applies to the divergence of $\widetilde{G}_{(0i)}$, which gives
\begin{eqnarray}
	\partial^i \widetilde{G}_{(0i)} &=& - 2\partial_t \partial^i \partial_i \psi - \frac{\gamma_W}{2}  \left[ \partial_i\partial^i \partial_t (\Phi -\psi+ \partial_t A/2) + \frac{1}{2} \partial^i \partial_i \square A \right] = 8\pi \widetilde{G}_N \partial^i \partial_i b\,,
\end{eqnarray}
so that
\begin{eqnarray}
	\partial_t \psi + ({\gamma_W}/{4}) \left[ \partial_t (\Phi -\psi+ \partial_t A/2) + \square A/2 \right] = -4\pi \widetilde{G}_N b\,. \label{EqG0ia}
\end{eqnarray}
Combining Eqs.~\eqref{EqG0ia} and \eqref{EqH0iS}, we can have
\begin{eqnarray}\label{EqHG1}
	\partial_t(2\psi+ \gamma_W\Phi) = 8\pi \widetilde{G}_N (s-b)\,.
\end{eqnarray}

Next we turn to Eq.~\eqref{EqGij}. We have two ways to extract equations only for scalar fields. One is to take the trace of this equation, which gives
\begin{eqnarray}\label{EqGii}
	\widetilde{G}^i_i &=& \frac{1}{2} \left\{2\partial^i \partial_i [(\gamma_W-1)\psi + \Phi -(\gamma_W+1)\partial_t A/2]+ 3 [(4-2\gamma_W)\square \psi - 2\partial_k \partial^k (\Phi+\psi-\partial_t A/2)]\right\} \nonumber\\
	&=& -8\pi \widetilde{G}_N (3p+ 2\partial^i \partial_j q)\,,
\end{eqnarray}
while the other way is to take the double divergences 
\begin{eqnarray}
	\partial^i \partial^j \widetilde{G}_{(ij)} &=& \frac{1}{2} \left\{ 2\partial^i \partial_j \partial^j \partial_j [(\gamma_W-1)\psi+\Phi-(\gamma_W+1)\partial_t A/2] \right. \nonumber\\
	&& \left.+ \partial_i\partial^i [(4-2\gamma_W)\square \psi- 2\partial^k\partial_k [\Phi+\psi-\partial_t A/2]] \right\} \nonumber\\
	&=& -8\pi \widetilde{G}_N (\partial^i \partial_j p + 2 \partial^j \partial_j \partial^i \partial_i q)\,,
\end{eqnarray}
which can be simplified into
\begin{eqnarray}\label{EqGija}
	&&\partial^j \partial_j [(\gamma_W-1)\psi+\Phi-(\gamma_W+1)\partial_t A/2] + \left[ (2-\gamma_W)\square \psi- \partial^k \partial_k (\Phi+\psi-\partial_tA/2) \right] \nonumber\\
	&=& -8\pi \widetilde{G}_N (p+ 2\partial_j \partial^j q)\,.
\end{eqnarray}
Combining Eqs.~\eqref{EqGija} and \eqref{EqGii} can lead us to the following two equations
\begin{eqnarray}
	&&(2-\gamma_W)\square\psi - \partial^k \partial_k (\Phi + \psi-\partial_t A/2) = -8\pi \widetilde{G}_N p\,, \label{EqGijb}\\
	&&(\gamma_W-1)\psi + \Phi -(\gamma_W+1) \partial_t A/2 = -16\pi \widetilde{G}_N q\,. \label{EqGijc} 
\end{eqnarray}
Further, by simultaneously taking into account Eqs.~\eqref{EqH0iS} and \eqref{EqGijb}, we can yield 
\begin{eqnarray}
	\square [(1-\gamma_W)  \psi- \Phi] = -8\pi\widetilde{G}_N (p-2 \partial_t s/\gamma_W)\,.\label{EqHG2}
\end{eqnarray}

Besides Eqs.~\eqref{EqHG1} and \eqref{EqHG2}, there are also other relations in Eq.~\eqref{EqG00}, \eqref{EqGijc}, \eqref{EqCiS} and \eqref{EqC0}. From the perspective of the dynamics, they are redundant. But they can give us constraints on the source fields. By integrating Eq.~\eqref{EqG00}, we can obtain 
\begin{eqnarray}\label{SolS}
	2\psi + \gamma_W \Phi = -\frac{2\widetilde{G}_N M}{r}\,
\end{eqnarray}
via the Green's function technique~\cite{Carroll:2004st}. Note that the right-hand side 
is constant in time, which means that $\partial_t (2\psi+ \gamma_W \Phi) = 0$ outside of the source material region. Also, we can combine Eqs.~\eqref{EqG00} and \eqref{EqHG1} to obtain
\begin{eqnarray}\label{SCons0}
	\square (2\psi+\gamma_W \Phi) = 8\pi \widetilde{G}_N (\rho + \partial_t b -\partial_t s)\,.
\end{eqnarray}
Thus, the far-field solution to this equation is given by
\begin{eqnarray}
	2\psi + \gamma_W\Phi = -\frac{2\widetilde{G}_N}{r} \int_{\cal V} d^3y \left[ \rho + \partial_{{t}} (b-s) \right] (\tilde{t}_r,\mathbf{y})\,,
\end{eqnarray}
in which $\tilde{t}_r = t-|\mathbf{r}-\mathbf{y}|$ is the retarded time with $\mathbf{r}$ and $\mathbf{y}$ denoting the spatial coordinates of the center of  mass and points inside the matter source, respectively. Here we have only kept terms up to the leading order of $1/r$ in the far-field expansion, which also implies $\tilde{t}_r\approx t-r + \hat{\mathbf{r}}\cdot \mathbf{y}$ with $\hat{\mathbf{r}}$ the unit vector along the $\mathbf{r}$ direction~\cite{Will:1977wq}. By comparing with the solution in Eq.~\eqref{SolS},  we can obtain the following relation
\begin{eqnarray}\label{ScalarCons}
	M = \int_{\cal V} d^3 y \left[\rho + \partial_t (b-s)\right](\tilde{t}_r, \mathbf{y})\,.
\end{eqnarray}
In Appendix~\ref{ProofScalar}, we show the exactness of this equality in the far-field limit~\cite{Will:1993hxu,Misner:1973prb,Will:1977wq}.  

By combining Eqs.~\eqref{EqHG2} and \eqref{SCons0}, we can obtain the following wave equation for $\psi$,
\begin{eqnarray}
	(1-\gamma_W/2) (1+\gamma_W) \square \psi = 4\pi \widetilde{G}_N \left[\rho - \gamma_W p + \partial_t (b+s)\right]\,.
\end{eqnarray}
The far-region solution to this equation is given by
\begin{eqnarray}
	 (1-\gamma_W/2)(1+\gamma_W) \psi &=& -\frac{\widetilde{G}_N}{r} \int_{\cal V} d^3 y \left[\rho - \gamma_W p + \partial_t (b+s)\right] (\tilde{t}_r,\mathbf{y}) \nonumber\\
	&=& -\frac{\widetilde{G}_N M}{r} + \frac{\widetilde{G}_N}{r} \int_{\cal V} d^3 y \left[\gamma_W p -2\partial_t s\right] (\tilde{t}_r ,\mathbf{y}) \nonumber\\
	&\approx&  -\frac{\widetilde{G}_N M}{r} + \frac{\widetilde{G}_N}{r} \int_{\cal V} d^3 y \left[\gamma_W p -2\partial_t s\right] (t_r,\mathbf{y})
\end{eqnarray}
where we have considered Eq.~\eqref{ScalarCons} in the second equality and have kept only the leading-order contribution in the long-wavelength expansion in the last approximation. 

We can represent the two volume integrals in the above equation with more physical variables. Let us firstly consider the following integral
\begin{eqnarray}\label{IntS}
	&&\int_{\cal V} d^3y s (t, \mathbf{y}) = \int_{\cal V} d^3 y \frac{1}{3} (\partial_i y^i) s(t, \mathbf{y}) = -\frac{1}{3}\int_{\cal V} d^3 y y^i \partial_i s(t, \mathbf{y}) \nonumber\\
	&=& \frac{1}{3} \int_{\cal V} d^3y y^i \left( T_{[0i]} - u_i \right) = \frac{1}{3} \int_{\cal V} d^3 y y^i T_{[0i]} \equiv \frac{1}{3}\widetilde{S}\,, 
\end{eqnarray}
and, according to Eq.~\eqref{EqCs}, we have
\begin{eqnarray}\label{IntP}
	&& \int_{\cal V} d^3 y p(t, \mathbf{y}) = -\int_{\cal V} d^3 y \left[ \partial_t (b+s) + 2 \partial^j \partial_j q  \right] = -\frac{d}{dt}\int_{\cal V} d^3y \frac{1}{3} \partial_j y^j (b+s) \nonumber\\
	&=& \frac{1}{3} \frac{d}{dt} \int_{\cal V}d^3y y^j \partial_j (b+s)  = \frac{1}{3} \frac{d}{dt} \int_{\cal V} d^3y y^j \left[(v_j+u_j) -T_{0j}\right] = - \frac{1}{3} \frac{d}{dt}  \int_{\cal V} d^3 y y^j (T_{j0} + 2 T_{[0j]})  \nonumber\\
	 &=&  \frac{1}{6}\frac{d^2}{dt^2}\int_{\cal V} d^3 y y^j y_j \rho - \frac{2}{3} \frac{d}{dt} \int_{\cal V} d^3y y^j T_{[0j]} =  \frac{1}{6} \frac{d^2}{dt^2}{\cal I}^k_{\,k} - \frac{2}{3} \frac{d}{dt} \widetilde{S}\,. 
\end{eqnarray}
where we have used the relations
\begin{eqnarray}
  \int_{\cal V} d^3y y^i  T_{i0} = -\frac{1}{2} \frac{d}{dt}\int_{\cal V} d^3 y y^j y_j \rho\,,\quad \int_{\cal V} d^3 y y^i u_i = 0\,,
\end{eqnarray}
and defined the following quantities
\begin{eqnarray}
	\widetilde{S} \equiv \int_{\cal V} d^3 y y^i T_{[0i]}\,, \quad \mathcal{I}^k_{\,k} = \int_{\cal V} d^3 y y^k y_k \rho \,.
\end{eqnarray}
Note that $\mathcal{I}^k_{\,k} = \delta^{ik} \mathcal{I}_{ik}$ is the trace of the matter quadrupole moment $\mathcal{I}_{ij}$ and thus explains this notation. Therefore, the solution of the field $\psi$ in the far region is given by
\begin{eqnarray}\label{SolPsi}
	\psi (t, \mathbf{x}) = -\frac{1}{1-\gamma_W} \frac{G_N M}{r}  + \frac{G_N}{(1-\gamma_W)r} \left[\frac{\gamma_W}{6}\frac{d^2}{dt^2_{r}} {\cal I}^k_{\,k} - \frac{2}{3}(1+\gamma_W) \frac{d}{dt_r} \widetilde{S}\right]\,,
\end{eqnarray}  
where we have used Eq.~\eqref{DefGN} to represent the result with the physical measured Newtonian constant $G_N$. By further using the constraint in Eq.~\eqref{SolS}, we can solve the field $\Phi$ as follows
\begin{eqnarray}\label{SolPhi}
	\Phi (t, \mathbf{x}) = -\frac{G_N M}{r} -\frac{G_N}{(1-\gamma_W) r} \left[\frac{1}{6} \frac{d^2}{dt_r^2} \mathcal{I}^k_{\,k} -\frac{2(1+\gamma_W)}{3\gamma_W} \frac{d}{dt_r} \widetilde{S} \right]\,. 
\end{eqnarray}
It is obvious that there are two distinct contributions to the scalar wave production: $\mathcal{I}^k_{\,k}$ and $\widetilde{S}$, where the former source comes from the symmetric part of the stress-energy tensor $T_{(\mu\nu)}$ while the latter from the antisymmetric part $T_{[\mu\nu]}$. This is contrasted with the Brans-Dicke gravitational theory~\cite{Brans:1961sx} in which only the symmetric EM tensor source $\mathcal{I}^k_{\,k}$ can contribute to the scalar GW radiation. Moreover, Eqs.~\eqref{SolPsi} and \eqref{SolPhi} show that, compared with the term related to $\mathcal{I}^k_{\,k}$, the contribution to scalar GWs from $\widetilde{S}$ is enhanced by a factor of $1/\gamma_W$ given the smallness of $\gamma_W\ll 1$. Interestingly, for the scalar $\Phi(t,\mathbf{x})$, the true coupling of scalar GWs to the source $\widetilde{S}$ can be further reduced in the GQFT as follows
\begin{eqnarray}
	\frac{G_N (1+\gamma_W)}{\gamma_W(1-\gamma_W)} = \frac{\widetilde{G}_N}{\gamma_W(1-\gamma_W/2)} \approx \frac{1}{8\pi m_G^2 (\alpha_G-\alpha_W/2)} \sim \frac{1}{8\pi m_G^2}\,,
\end{eqnarray}
where we have used the relation in Eq.~\eqref{DefGN} and the following basic definitions~\cite{Gao:2024juf,Wu:2022mzr}
\begin{eqnarray}
	\gamma_W\equiv \gamma_G (\alpha_G-\alpha_W/2)\,,\quad \gamma_G\equiv 8\pi \widetilde{G}_N m_G^2\,.
\end{eqnarray}
Here $\alpha_{G,W}$ are two $\mathcal{O}(1)$ dimensionless parameters while $m_G$ is the another fundamental scale characterizing the typical masses of spin gauge fields obtained by breaking the scale symmetry in the GQFT. Given the strong upper bound on $\gamma_W < \mathcal{O}(10^{-5})$  by local gravity tests in the Solar System~\cite{Gao:2024juf}, we expect that this constraint can be transformed to $m_G\lesssim 10^{16}~\mathrm{GeV}$. This indicates that the antisymmetric stress-energy tensor $T_{[\mu\nu]}$ plays a more significant role in the scalar GW generation with its coupling of order of $1/m_G^2$, while the contribution from symmetric part $T_{(\mu\nu)}$ would be suppressed by $G_N\sim 1/m_{\rm Pl}^2$ with $m_{\rm Pl}$ denoting the Planck scale.

On the other hand, for a fermion $\psi$ charged under some gauge symmetry groups, the precise definition of its antisymmetric EM tensor is given by~\cite{Wu:2022mzr}
\begin{eqnarray}
	T_{[\mu\nu]} &=& \frac{1}{4} \left(\chi^a_{\mu} \bar{\psi} \gamma_a i D_\nu \psi - \chi_\nu^a \bar{\psi} \gamma_a i D_\mu \psi + {\rm H.c.} \right) \nonumber\\
	&\approx& \frac{1}{4} \left( \bar{\psi} \gamma_\mu i D_\nu \psi - \bar{\psi} \gamma_\nu iD_\mu \psi + {\rm H.c.} \right)\,,  
\end{eqnarray}
where $D_\mu$ denotes the covariant derivative of gauge group acting on $\psi$. In the second line, we have taken the flat spacetime limit by approximating $\chi^a_\mu \approx \eta^a_\mu$. 
We can extract the longitudinal component $s$ and the associated scalar GW source $\widetilde{S}$ by explicitly computing $T_{[0i]}$. As shown in Appendices~\ref{uDef} and \ref{SDef}, when fermions are free or subject to a nontrivial conventional gauge field background in the flat spacetime limit, the spatial divergence of $T_{[0i]}$ always vanishes. Consequently, the scalar GW source $s$ or $\widetilde{S}$ cannot be generated under such an approximation. However, in the presence of a background spin gauge field $A_{\mu ab} \Sigma^{ab}$~\cite{Wu:2022mzr}, where $\Sigma^{ab}$ denotes the generators in the spinor representation of the spin gauge symmetry, Appendix~\ref{SDef} demonstrates that even in the flat spacetime limit, the spatial divergence $\partial^i T_{[0i]}$ can generally be nonzero.
Its explicit expression is given by: 
\begin{eqnarray}\label{pT0iA}
		\partial^i T_{[0i]} &=& \frac{g_G}{4}\left(\frac{i}{4} F_{\mu\nu}^{~~ab} \bar{\psi} \left[ \{\gamma_0, \Sigma^{\mu\nu}\}, \Sigma_{ab} \right]\psi  + m A_{\mu ab} \bar{\psi}\{\Sigma_0^{~\mu},\Sigma^{ab}\} \psi  \right. \nonumber\\
	&& + i A_{\nu ab} \left( \bar{\psi} \Sigma^{\mu\nu} \gamma_0 \Sigma^{ab} D_\mu \psi - D_\mu \bar{\psi} \Sigma^{ab} \gamma_0 \Sigma^{\mu\nu} \psi \right) \nonumber\\
	&& \left. + \frac{1}{2} A_{\mu ab} \left( \bar{\psi} \Sigma^{ab} \gamma_0 D^\mu \psi  + D^\mu \bar{\psi} \gamma_0 \Sigma^{ab} \psi - \bar{\psi} \Sigma^{ab} \gamma^\mu D_0 \psi - D_0 \bar{\psi} \gamma^\mu \Sigma^{ab} \psi \right)  \right)\,,
\end{eqnarray}
where the covariant derivative is defined as $D_\mu \psi = \partial_\mu - i g_G (A_{\mu ab}/2) \Sigma^{ab} \psi$ with $g_G$ being the corresponding gauge coupling. 

Thus, the scalar GW source $\widetilde{S}$ can be expressed as:
\begin{eqnarray}
	\widetilde{S} = - \frac{1}{2} \int_{\cal V} d^3y y^i y_i \partial^j T_{[0j]}\,,
\end{eqnarray}
where $\partial^j T_{[0j]}$ is given by Eq.~\eqref{pT0iA}. 

In GQFT, the spin gauge field $A_{\mu ab}$ is assumed to be massive, which motivates consideration of extreme situations at high energy scales, such as the early Universe or the interior of neutron stars (NSs). However, a detailed discussion of scalar GW production in these scenarios lies beyond the scope of this work. Instead, we now turn to another scalar GW source: the trace of the quadrupole moment $\mathcal{I}^k_k$. As an illustrative example, we consider a compact binary system with an elliptical orbit. For small orbital eccentricity $e$, we can even derive analytic expressions for the scalar fields $\psi$ and $\Phi$.

\subsubsection{Compact Object Binary with an Orbit of Small Ellipticity}

\begin{figure}[htb]
	\includegraphics[width=0.6\linewidth]{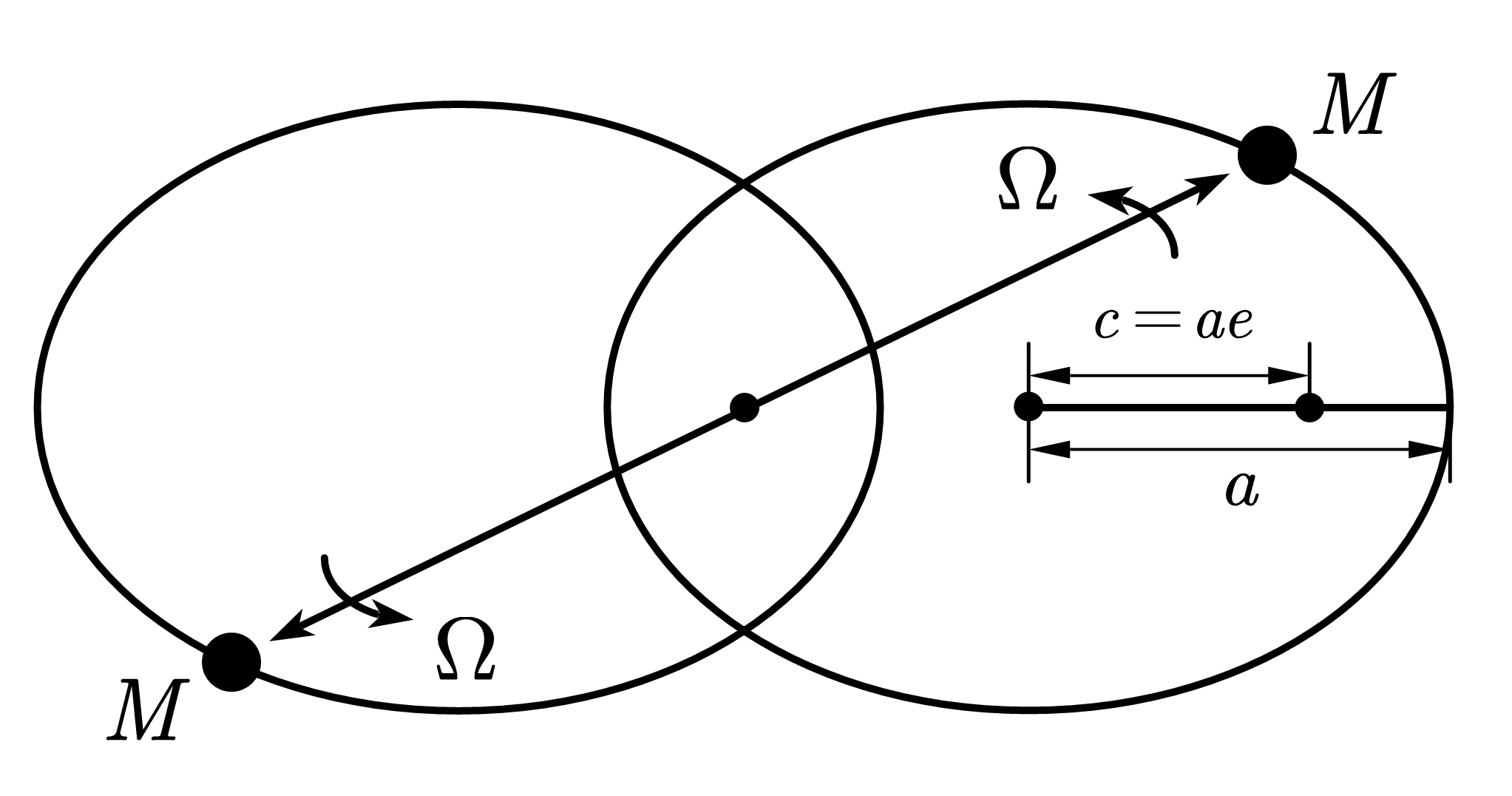}
	\caption{An illustration of the binary system with an elliptical orbit, which can generate scalar GWs.}
	\label{fig:elliptic}
\end{figure}

In the GQFT, in the absence of $\widetilde{S}$ from the anti-symmetric EM tensor,  the dominant source to the scalar GW mode is the second-order differentiation of $\mathcal{I}^k_{\,k}$ with respect to the retarded time $t_r$.  However, in the conventional binary system composed of compact objects, the orbit of each component is usually assumed to be a perfect circle, which leads to a constant $\mathcal{I}^k_{\,k}$ and the vanishing of its time derivatives. Thus, in this subsection, we would like to study a binary system with an elliptical orbit as illustrated in Fig.~\ref{fig:elliptic}, which is shown to offer a source to the non-zero $d^2 \mathcal{I}^k_{\,k} /dt_r^2$ and associated scalar GWs.  We shall consider the small ellipticity approximation for the binary orbit in order to simplify our analytic calculation. Also, we work in the Newtonian gravity which is already enough to exhibit the phenomenon. 

Let us consider an isolated binary in which two stars or black holes are exactly same with their respective mass $M$. Hence, the system satisfies the angular momentum conservation
\begin{eqnarray}\label{consL}
	L = 2 M r^2 \left( \frac{d\varphi}{dt}\right) \,,
\end{eqnarray}
and the energy conservation 
\begin{eqnarray}\label{consE}
	E = 2\times \left[\frac{M}{2} \left(\frac{dr}{dt}\right)^2 + \frac{L^2}{8Mr^2} \right] - \frac{G_N M^2}{2r}\,,
\end{eqnarray}
in which $r$ and $\varphi$ denote the distance and the azimuthal angle, respectively. As shown in Ref.~\cite{Gao:2024juf}, the Newton potential is exactly the same as in GR, since the GQFT meets the Weak Equivalence Principle for non-relativistic objects. By combining Eqs.~\eqref{consL} and \eqref{consE}, we can obtain the famous Binet equation~\cite{Will:2018bme}
\begin{eqnarray}
	\frac{d^2 u}{d\varphi^2} + u = \frac{1}{\bar{L}^2}\,,
\end{eqnarray}
where the variable is changed from $r$ to $u \equiv 1/r$ and we have defined the parameter $\bar{L}^2 \equiv L^2/(G_N M^3)$. The solution to the Binet equation is given by
\begin{eqnarray}\label{Orbit}
	u = \frac{1}{\bar{L}^2} (1+ e \cos\varphi)\,,\quad r = \frac{\bar{L}^2}{1+e \cos\varphi}\,,
\end{eqnarray}
where $e$ is the ellipticity of one star orbit. If $e<1$, the orbit is an ellipsis in which the focus is located at $r=0$ and the perihelion is directed in $\varphi =0$. 

To further proceed to obtain the orbit as a function of time $t$, one needs to solve the following differential equation transformed from Eq.~\eqref{consL}, 
\begin{eqnarray}\label{consL2}
	\frac{d\varphi}{dt} = \frac{L}{2Mr^2} = \frac{G^2 M^5}{2L^3} (1+e\cos\varphi)^2\,.
\end{eqnarray}
In order to simplify our following discussion, we would like to work in the small ellipticity limit, {\it i.e.}, $e\ll 1$, in which Eq.~\eqref{consL2} can be simplified into
\begin{eqnarray}
	\frac{G^2 M^5}{2L^3} dt = \frac{d\varphi}{(1+e\cos\varphi)^2} \approx d\varphi (1- 2e\cos\varphi)\,,
\end{eqnarray}
so that the approximate solution is given by
\begin{eqnarray}\label{PhiT}
	\varphi (t) \approx \Omega t + 2 e \sin\left(\Omega t\right)\,,
\end{eqnarray}
where $\Omega$ can be viewed as the mean angular frequency and is defined by
\begin{eqnarray}
	\Omega = \frac{G^2 M^5}{2L^3} = \frac{G^{1/2} M^{1/2}}{2a^{3/2}}\,.
\end{eqnarray}
By taking this solution into the elliptical orbit in Eq.~\eqref{Orbit}, we can also obtain the distance $r$ as a function of $t$ as follows
\begin{eqnarray}\label{RT}
	r \approx a \left[ 1 - e\cos\left(\Omega t\right) \right]\,,
\end{eqnarray} 
which is valid up to the leading order in $e$. Here we have defined the length of the major 
\begin{eqnarray}
	a = \frac{L^2}{ G M^3}\,.
\end{eqnarray}
This means that we can replace the angular momentum with the major $a$ in our following formulae. The period of the orbit can be easily yielded as follows
\begin{eqnarray}\label{Pe0}
	T = \frac{2\pi}{\Omega} = \frac{4\pi L^3}{G^{1/2} M^{1/2}} = \frac{4\pi a^{3/2}}{G^{1/2} M^{1/2}}\,.
\end{eqnarray}	
By using the fact that the velocity along the radial direction vanishes at the perihelion with $r_p = a(1-e)$ or at the aphelion with $r_a = a(1+e)$,  we yield the total energy of the system
\begin{eqnarray}\label{Ee0}
	E = \frac{L^2}{4mr_p^2} - \frac{GM^2}{2r_p} = -\frac{GM^2}{4a}\,. 
\end{eqnarray}
The results in Eqs.~\eqref{Pe0} and \eqref{Ee0} is in perfect agreement with the general non-perturbative solutions for elliptical orbits based on the Kepler's laws.  

Now we can show that this binary system with an elliptical orbit can produce a non-trivial trace of the quadrupole moment, which is defined by 
\begin{eqnarray}
	\mathcal{I}_{ij} (t) = \int_{\cal V} d^3  y\, y_i y_j \rho(t,\mathbf{y})\,.
\end{eqnarray}
For this binary system, the energy density can be written by
\begin{eqnarray} 
	&&T_{00} (t, \mathbf{y}) = \rho (t, \mathbf{y}) \nonumber\\
	&=& M \delta(y^3) \left[\delta\left(y^1-r\cos\varphi \right) \delta\left(y^2 - r\sin\varphi\right) + \delta\left(y^1+r\cos\varphi \right) \delta\left(y^2 + r\sin\varphi \right) \right] \,,
\end{eqnarray}
where $r$ and $\varphi$ evolve according to Eqs.~\eqref{PhiT} and \eqref{RT}, respectively. The trace of the quadrupole is given by
\begin{eqnarray}\label{Ss}
	\mathcal{I}^k_{\,k} = 2Mr^2(t) = 2Ma^2 (1-e\cos(\Omega t))^2 \approx 2 Ma^2 (1-2e\cos \Omega t)\,. 
\end{eqnarray}
As a result, in the GQFT, the scalar GW in the far region are given by
\begin{eqnarray}
	\psi (t, \mathbf{x}) &=& \frac{\gamma_W}{(1-\gamma_W/2)(1+\gamma_W)}  \frac{2\widetilde{G}_N M }{3r} (a\Omega)^2 e \cos(\Omega t_r)\nonumber\\
	&=&  \left(\frac{\gamma_W}{1-\gamma_W}\right) \frac{2G_N M}{3r} (a\Omega)^2 e \cos(\Omega t_r)\,, 
\end{eqnarray} 
where $t_r \equiv t-r = t-|\mathbf{r}|$, and we ignore the static potential part in Eq.~\eqref{SolPsi}.

\subsection{Vector Sector}
We now consider Eq.~\eqref{EqH0i}, which, after considering the scalar relation in Eq.~\eqref{EqH0iS}, can be simplified as follows
\begin{eqnarray}\label{EqH0iV}
	-({\gamma_W}/{4}) \left[ \square S_i - \partial_k \partial^k (S_i - \partial_t F_i) \right] = -8\pi \widetilde{G}_N u_i\,.
\end{eqnarray}
Also, by taking the associated scalar part in Eq.~\eqref{EqG0ia} into the symmetric part equation of  Eq.~\eqref{EqG0i}, we can yield the following relation
\begin{eqnarray}\label{EqG0iV}
	\frac{1}{2} \partial_k \partial^k (S_i -\partial_t F_i) + \frac{\gamma_W}{4} \left[ \square S_i + \partial_k \partial^k (S_i - \partial_t F_i) \right] = -8\pi \widetilde{G}_N v_i\,.
\end{eqnarray}
Adding the above two equations gives
\begin{eqnarray}
	\partial_k \partial^k (S_i - \partial_t F_i) = -\frac{16\pi \widetilde{G}_N}{1+\gamma_W} (v_i + u_i)\,,
	\end{eqnarray}
from which, by using the Green's function technique~\cite{Carroll:2004st}, we can obtain
\begin{eqnarray}\label{EqHGV1}
	&& S_i - \partial_t F_i = \frac{4\widetilde{G}_N}{(1+\gamma_W)}\int_{\cal V} d^3y \frac{v_i+u_i}{|\mathbf{r}-\mathbf{y}|} \approx \frac{4 \widetilde{G}_N}{1+\gamma_W} \left[\frac{1}{r} \int_{\cal V} d^3y (v_i+u_i) + \frac{\hat{k}^j}{r^2} \int_{\cal V} d^3y y_j (v_i+u_i)\right]\nonumber\\
	 &= & \frac{4\widetilde{G}_N}{(1+\gamma_W)} \left[ \frac{K_i}{r} - \frac{1}{2r^2}\epsilon_{ijk} \hat{k}^j L^k\right] = -\frac{2\widetilde{G}_N}{(1+\gamma_W)} \frac{\epsilon_{ijk} \hat{k}^j L^k}{r^2}\,,
\end{eqnarray} 
in which $\hat{k}_j$ denotes the GW propagation direction, which lies along with $\mathbf{r}$, {\it i.e.}, $\hat{k}_j = \hat{r}_j $.  $K_i$ is defined as in Eq.~\eqref{DefMK} and can be taken to be zero as in Eq.~\eqref{ProofMK} and \eqref{ProofK0}, while $L^k$ is the conserved angular momentum defined in Eq.~\eqref{DefL}. In the derivation, we have used the following identity
\begin{eqnarray}\label{idV2}
	&&\int_{\cal V} d^3 y \left[y_i (v_j +u_j ) + y_j (v_i+u_i)\right] \nonumber\\
	&=& \int_{\cal V} d^3y \partial^k \left[y_i y_j (v_k + u_k)\right] - \int_{\cal V} d^3 y y_j y_j \partial^k (v_k +u_k) =0\,,
\end{eqnarray}
where the first term vanishes since it is a total divergence, while the second term is zero due to the transverse property of $u_i$ and $v_i$. Eq.~\eqref{EqHGV1} can be viewed as a constraint imposed on the dynamics of the vector GW fields $S_i$ and $F_i$. 

By taking the divergence of Eq.~\eqref{EqGij},  one can derive the following equation
\begin{eqnarray}
	\partial^j G_{ij} = (1/4) \left[ \gamma_W \square \partial_j \partial^j F_i + (\gamma_W+2) \partial_t \partial^j \partial_j (S_i - \partial_t F_i) \right] = -8\pi\widetilde{G}_N \partial^j \partial_j \theta_i\,,
\end{eqnarray}
which can be reduced to
\begin{eqnarray}\label{EqFVW}
	(1/4) \left[ \gamma_W \square F_i + (\gamma_W +2) \partial_t (S_i -\partial_t F_i) \right] = -{8\pi \widetilde{G}_N}\theta_i\,.
\end{eqnarray}
This equation, together with Eq.~\eqref{EqG0iV}, can give us 
\begin{eqnarray}
	\square (S_i - \partial_t F_i) = -16\pi \widetilde{G}_N (v_i- \partial_t\theta_i) /(1+\gamma_W)\,.
\end{eqnarray}
Thus, the solution to this equation gives another form of $S_i - \partial_t F_i$ as
\begin{eqnarray}\label{EqHGV2}
	&& S_i - \partial_t F_i = \frac{4\widetilde{G}_N}{1+\gamma_W} \int_{\cal V} d^3 y \frac{(v_i -\partial_{\tilde{t}_r}\theta_i) (\tilde{t}_r, \mathbf{y})}{|\mathbf{x}-\mathbf{y}|} \nonumber\\
	&\approx&  \frac{4\widetilde{G}_N}{1+\gamma_W} \left[ \frac{1}{r}\int_{\cal V} d^3y (v_i -\partial_{\tilde{t}_r} \theta_i) + \frac{\hat{k}_j}{r^2} \int_{\cal V} d^3 y y^j (v_i -\partial_{\tilde{t}_r} \theta_i) \right] \,,
\end{eqnarray}
where we have kept terms up to $\mathcal{O}(1/r^2)$ in the far-field expansion. Here the vector fields $v_i$ and $\theta_i$ are functions of retarded time $\tilde{t}_r \equiv t- |\mathbf{r}-\mathbf{y}|\approx t-r + \hat{\mathbf{k}}\cdot \mathbf{y}$, which provides an extra dependence on integration coordinates $\mathbf{y}$. Since Eqs.~\eqref{EqHGV1} and \eqref{EqHGV2} give expressions of the same quantity, the right-hand side of both equations should be the same for consistency. 

On the other hand, our focus here is the far-field solution, which should contain two components. One component is GW radiations, for which only the terms of $\mathcal{O}(1/r)$ should be of relevance. This is due to the fact that the GW energy flux would be the square of GW fields, and in order to propagate to the far region, the GW energy flux should decay as $1/r^2$ at most. Otherwise, the energy flux decays too fast to reach a distant observatory. The second component is related to the conserved charges in the system, which can appear in terms either of $\mathcal{O}(1/r)$ or of $\mathcal{O}(1/r^2)$ since only those two terms can give nonzero surface integrals at spatial infinity. Therefore, in comparison of Eqs.~\eqref{EqHGV1} and \eqref{EqHGV2}, we should have the following two relations
\begin{eqnarray}\label{ConsistencyV}
	\int_{\cal V} d^3 y (v_i - \partial_{\tilde{t}_r} \theta_i) (\tilde{t}_r, \mathbf{y}) = 0\,,\quad  \hat{k}^j \int_{\cal V} d^3 y y_j (v_i -\partial_t \theta_i) \simeq -\frac{1}{2}\epsilon_{ijk} \hat{k}^j L^k \,.
\end{eqnarray}
Here we have used in the second relation the notation $\simeq$ to notify that it is valid only up to the leading order in the long-wavelength expansion with respect to $(\hat{\mathbf{k}}\cdot\mathbf{y})\partial_{t_r}$. We can directly prove these consistency conditions with the conservation relations and the long-wavelength expansion technique~\cite{Will:1977wq}, which are detailed in Appendix~\ref{ProofVector}.



We now turn back to solve the vector GW radiation equations in the GQFT. With the relations in Eq.~\eqref{EqH0iV} and \eqref{EqG0iV}, we can also obtain
\begin{eqnarray}
	\square S_i = 16\pi \widetilde{G}_N \left[ \frac{2}{\gamma_W} u_i - \frac{1}{1+\gamma_W} (u_i + v_i) \right]\,.
\end{eqnarray}
By using the Green's function, one can derive
\begin{eqnarray}\label{GreenV}
	S_i(t, {\bf r}) = -{4\widetilde{G}_N} \int_{\cal V} d^3 y \frac{1}{|\mathbf{r}-\mathbf{y}|} \left[ \frac{2}{\gamma_W}  u_i (\tilde{t}_r, {\bf y}) - \frac{1}{1+\gamma_W} (v_i + u_i) (\tilde{t}_r, \mathbf{y}) \right]\,.
\end{eqnarray}
Since we are interested in the radiation solution, we can expand the factor $1/|\mathbf{x}-\mathbf{y}| \approx 1/r + \hat{k}^j y_j /r^2 + \mathcal{O}(1/r^3)$ and, for the reason detailed in the previous paragraph, only keep the leading-order contribution as follows
\begin{eqnarray}
	S_i (t, \mathbf{r}) \approx -\frac{4\widetilde{G}_N}{r} \int_{\cal V} d^3y \left[ \frac{2}{\gamma_W}  u_i (\tilde{t}_r, {\bf y}) - \frac{1}{1+\gamma_W} (v_i + u_i) (\tilde{t}_r, \mathbf{y}) \right]\,.
\end{eqnarray}
We further assume that the radiated GW wavelength is much longer than the typical length scale of source region, which means that $(\hat{\mathbf{k}}\cdot\mathbf{y})\partial_{t_r} \ll 1$. Hence, we can expand $S_i$ further as follows
\begin{eqnarray}
	S_i (t, \mathbf{y}) \approx -\frac{4 \widetilde{G}_N}{r}  \sum_{n=0}^\infty \frac{1}{n!} \hat{k}_{i_1}  \cdots \hat{k}_{i_n} \int_{\cal V} d^3 y y^{i_1}  \cdots y^{i_n} \partial_{t_r}^n  \left[ \frac{2}{\gamma_W}  u_i ({t}_r, {\bf y}) - \frac{1}{1+\gamma_W} (v_i + u_i) ({t}_r, \mathbf{y}) \right]\,. \nonumber\\
\end{eqnarray}
We shall compute the leading-order terms in this series
\begin{itemize}
	\item $n=0$:  
	\begin{eqnarray}
		S_i^{(0)} = -\frac{4\widetilde{G}_N}{r}\int_{\cal V} d^3 y \left[\frac{2}{\gamma_W} u_i - \frac{1}{1+\gamma_W} (v_i +u_i) \right] =0\,,
	\end{eqnarray}
	due to the transverse nature of the $v_i$ and $u_i$. 
	\item $n=1$:
	\begin{eqnarray}\label{SolV1}
		&& S_i^{(1)} = -\frac{4\widetilde{G}_N}{r} \hat{k}^{j} \frac{d}{d t_{r}} \int_{\cal V} d^3y y_j \left[\frac{2}{\gamma_W} u_i -\frac{1}{1+\gamma_W} (v_i +u_i)\right] \nonumber\\
		&=& \frac{2\widetilde{G}_N}{r} \epsilon_{ijk} \hat{k}^j \frac{d}{dt_r}  \left\{ \frac{2}{\gamma_W} U^k  - \frac{1}{1+\gamma_W} L^k \right\} = \frac{4}{\gamma_W}\frac{\widetilde{G}_N}{r} \epsilon_{ijk} \hat{k}^j \frac{d}{dt_r} U^k \,,
	\end{eqnarray}
	where $L^k$ is the angular momentum defined in Eq.~\eqref{DefL} while $U^k$ is defined by
	\begin{eqnarray}
		U^k = \int_{\cal V} d^3y \epsilon^{kji} y_j u_i \,.
	\end{eqnarray}
	Note that we have found in Eq.~\eqref{ConstL} that the angular momentum $L^k$ is conserved with $dL^k/dt=0$, so that the leading-order contribution to producing the vector GW mode is from the antisymmetric EM tensor components $u_i$. 
	\item $n\geqslant 2$: These terms are suppressed compared with the contribution $S_i^{(1)}$ in the long-wavelength limit, and can thus be neglected.
\end{itemize}

As illustrated in Appendix~\ref{uDef}, for the physical sources composed of non-relativistic fermion particles, the vector field $u_i$ in the antisymmetric EM tensor should be related to the fermionic spin density $\Xi^k(\mathbf{x})$ via
\begin{eqnarray}
	u_i = -\frac{1}{2} \epsilon_{ijk} \partial^j \Xi^k(t, \mathbf{x})\,.
\end{eqnarray}
As a result, the quantity $U^k$ can be easily calculated as 
\begin{eqnarray}
	U^k (t) = -\int_{\cal V} d^3 y \Xi^k (t, \mathbf{y})\,,
\end{eqnarray} 
which means that $U^k$ can be straightforwardly interpreted as minus the total spin of the fermionic matter. Hence, the vector GWs in the GQFT are predominantly generated by the time variation of the net total spin of matter fermions.  

In the above derivation, it is seen that the far-away radiation solely comes from $u_i$ in Eq.~\eqref{GreenV} without any dependence on  $(v_i + u_i)$. However, when expanding $1/|\mathbf{r}-\mathbf{y}|$ to the order of $1/r^2$, the term related to $(v_i+u_i)$ can give rise to the following extra contribution associated with the conserved charge $L^i$
\begin{eqnarray}
	S_i^{(L)} = -\frac{2\widetilde{G}_N}{(1+\gamma_W)} \frac{\epsilon_{ijk}\hat{k}^j L^k}{r^2}\,,
\end{eqnarray}
where we have only kept the leading-order result in the small $\mathbf{y}$ expansion. Note that at the same $1/r^2$ order, the field $u_k$ can also contribute as shown in Eq.~\eqref{GreenV}, but this term is much smaller than that given in Eq.~\eqref{SolV1} in the large $r$ limit, and can thus be ignored. As a result, the dominant vector field far away from the source is given by
\begin{eqnarray}\label{SolS1}
	S_i = S_i^{(L)} + S_i^{(1)} = -\frac{2(1-\gamma_W/2) G_N}{(1-\gamma_W)} \left[\frac{\epsilon_{ijk}\hat{k}^j L^k}{r^2} -\frac{2(1+\gamma_W)}{\gamma_W} \frac{\epsilon_{ijk}\hat{k}^j (dU^k/dt_r)}{r} \right] \,,
\end{eqnarray} 
where we have also expressed the final result in terms of the physical Newtonian constant $G_N$ by applying the identity of Eq.~\eqref{DefGN}. By further making use of Eq.~\eqref{EqHGV1}, the other vector field $F_i$ in the theory can be obtained as 
\begin{eqnarray}\label{SolF}
	F_i = \frac{4 (1-\gamma_W/2)(1+\gamma_W)}{\gamma_W(1-\gamma_W)} \frac{G_N}{r} \epsilon_{ijk}\hat{k}^j U^k\,.
\end{eqnarray}
Note that the general solutions $S_i$ and $F_i$ are automatically transverse to the propagating direction as $k^i S_i = k^i F_i =0$, so that we do not need to impose any further constraint.

It is manifest that the vector gravitational radiation comes only from $T_{[\mu\nu]}$, in which the corresponding coupling can be estimated by
\begin{eqnarray}
	\frac{(1-\gamma_W/2)(1+\gamma_W) G_N}{\gamma_W(1-\gamma_W)} = \frac{\widetilde{G}}{\gamma_W} = \frac{1}{8\pi (\alpha_G-\alpha_W/2)m_G^2}\approx \frac{1}{8\pi m_G^2}\,,
\end{eqnarray}
with $\alpha_{G(W)}$ taken to be $\mathcal{O}(1)$ in the last approximation. Therefore, similar to the antisymmetric EM contribution $\widetilde{S}$ to the scalar GW production, the vector GW production should be enhanced compared with the usual tensor counterpart.

\begin{figure}[t]
	\includegraphics[width=0.6\linewidth]{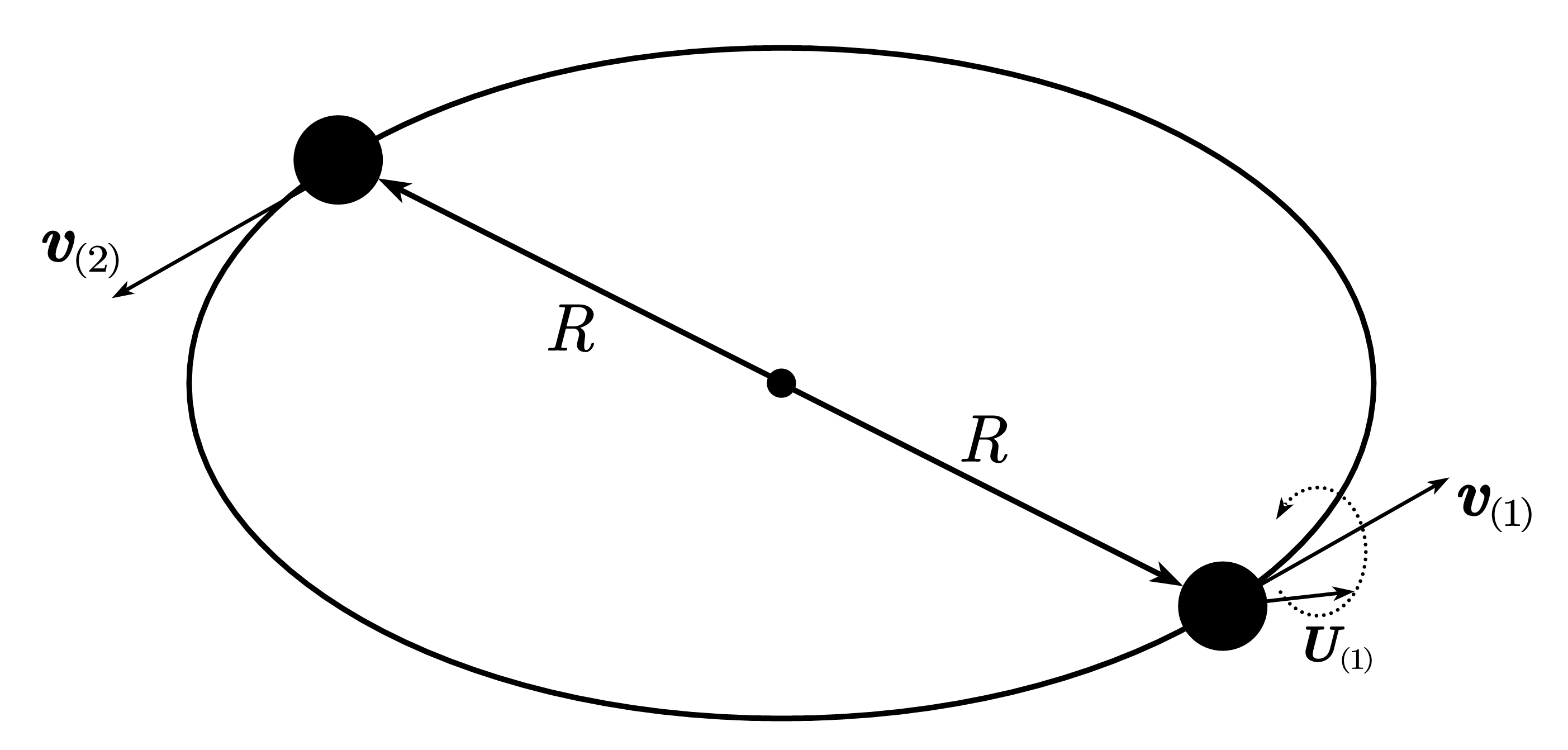}
	\caption{An illustration of a NS binary in which the spin of one star changes with its velocity.}
	\label{fig:spin}
\end{figure}

As an application of the above formalism, we shall consider a NS binary in which each star turning around one another in a circular orbit. For simplicity, these two NSs are assumed to share the same mass $M$. However, the total spin $-U_1^i$ of one NS composed of all or a substantial fraction of particle spins is nonzero and always aligns its orientation with the star's velocity, while the net spin of its companion star vanishes by averaging fermion spins contained. The system is displayed in Fig.~\ref{fig:spin}. As a result, we can write down, in the center-of-mass frame, the orbit of the spinning NS
\begin{eqnarray}
	\mathbf{R}_{(1)}  =  (R\cos (\omega t+\varphi_0), R\sin (\omega t+\varphi_0), 0) \,,
\end{eqnarray}
where $R$, $\omega$ and $\varphi_0$ refer to the orbit radius, angular frequency, and the initial phase at $t=0$, respectively. With the Kepler's third law, the angular frequency can be yielded as 
\begin{eqnarray}
       \omega = \sqrt{G_N M/(4 R^3)}.
\end{eqnarray}
Thus, the velocity of the star is $\mathbf{v}_{(1)} = R\omega \hat{v}$ where the unit vector $\hat{v} = (-\sin(\omega t + \varphi_0), \cos(\omega t+ \varphi_0), 0)$ denotes the velocity direction. By assuming that the spin of this NS changes with its velocity, we can express it as $-U^i_{(1)} =  U_0 \hat{v}^i (t)$, where $U_0$ is the constant amplitude of the spin vector. By taking this time-dependent NS spin into Eqs.~\eqref{SolS1} and \eqref{SolF}, we can obtain the following vector GW waveform 
\begin{eqnarray}
	S_i &=& \frac{(1-\gamma_W/2)(1+\gamma_W)}{\gamma_W(1-\gamma_W)} \frac{4G_N \omega U_0}{Rr} \epsilon_{ijk} \hat{k}^j {{R}^k_{(1)}} (t_r)\,,\nonumber\\
	F_i &=& -\frac{(1-\gamma_W/2)(1+\gamma_W)}{\gamma_W(1-\gamma_W)} \frac{4 G_N U_0}{r} \epsilon_{ijk} \hat{k}^j \hat{v}^k(t_r)\,,
\end{eqnarray}
where we have ignored the static part in Eq.~\eqref{SolS1} in the zone far away from the source.  



\subsection{Tensor Sector}
After investigating the scalar and vector sectors, we now consider the tensor GW generation in the GQFT. By taking into account all the relations for scalar and vector fields, the tensor GW equation is given by
\begin{eqnarray}	
	\square \hat{h}_{ij} = -\frac{16\pi \widetilde{G}_N}{1+\gamma_W} \sigma_{ij} (t, {\bf x})\,.
\end{eqnarray}
With the Green's function techniques, we can derive the following general solution
\begin{eqnarray}
	\hat{h}_{ij} \approx \frac{4\widetilde{G}_N}{(1+\gamma_W)r} \int_{\cal V} d^3 y \sigma_{ij} ({t}_r, {\bf y})\,,
\end{eqnarray}
where we have only remained the leading-order contribution in the far-field expansions. 
Note that $\sigma_{ij}$ is a symmetric tensor, {\it i.e.}, $\sigma_{ij} = \sigma_{ji}$, so that we can make use of this property to simplify the expression. 
\begin{eqnarray}\label{IntSig}
	\sigma_{ij} = (1/2) (T_{ij} + T_{ji}) -p\delta_{ij} - 2 \partial_{(i} \theta_{j)} -2 \partial_i \partial_j q\,,
\end{eqnarray}
where $T_{ij} = T_{(ij)}+ T_{[ij]}$ is the total stress-energy tensor. Thus, the integration over the compact source region is given by 
\begin{eqnarray}
	2 \int_{\cal V} d^3y \sigma_{ij} = \int_{\cal V} d^3y \left[ (T_{ij} + T_{ji}) -2 p\delta_{ij} \right]\,.
\end{eqnarray}
Recall the conservation conditions in the GQFT~\cite{Carroll:2004st}
\begin{eqnarray}
	\partial^\mu T_{\mu\nu} = 0\,,
\end{eqnarray}
which leads to the following identity
\begin{eqnarray}
	&&\int_{\cal V} d^3 y T_{ij} = \int_{\cal V} d^3y (\partial^k y_i) T_{kj} = -\int_{\cal V} d^3 y y_i \partial^k T_{kj} = -\int_{\cal V} d^3 y y_i \partial_t T_{0j} \nonumber\\
	&=& -\frac{d}{dt} \int_{\cal V} d^3y y_i (T_{j0}+2T_{[0j]}) = -\frac{d}{dt} \int_{\cal V} d^3y y_i (T_{j0}+2u_j- 2\partial_j s)\,.
\end{eqnarray}
Hence, we can simplify Eq.~\eqref{IntSig} as follows
\begin{eqnarray}
2\int_{\cal V} d^3y \sigma_{ij} &=& - \frac{d}{dt} \int_{\cal V} d^3y \left[ (y_i T_{j0} +y_j T_{i0}) +2 (y_i u_j+ y_j u_i) - 2(y_i\partial_j s + y_j \partial_i s) \right] - \int_{\cal V} d^3y (2p)\delta_{ij} \nonumber\\
	&=& -\frac{d}{dt} \int_{\cal V} d^3y \left[ \partial^l (y_i y_j T_{l0}) - y_i y_j\partial^lT_{l0}\right] -\frac{d}{dt} \int_{\cal V} 2\left[ \partial^l(y_i y_j u_l) - y_j y_j \partial^l u_l \right] \nonumber\\
	&& - \frac{d}{dt} \int_{\cal V} d^3 y 2\left[ (\partial_jy_i + \partial_i y_j)s \right]  - \int_{\cal V} d^3y 2p\delta_{ij} \nonumber\\
	&=& \frac{d}{dt} \int_{\cal V} d^3y y_i y_j \partial_t \rho  - \int_{\cal V} d^3y 2 (p+ 2\partial_t s) \delta_{ij} = \frac{d^2}{dt^2} \left({\cal I}_{ij}-\frac{1}{3} \delta_{ij}\mathcal{I}^k_{\,k}\right)\,,
\end{eqnarray}
where we have defined the quadruple tensor
\begin{eqnarray}
	{\cal I}_{ij} \equiv \int_{\cal V} d^3 y y_i y_j \rho\,,
\end{eqnarray} 
and used the trasversality of the vector $\partial^i u_i = 0$. We have also taken into account the identities in Eqs.~\eqref{IntS} and \eqref{IntP} to simplify the final expression. Finally,  given the tensorial GWs should be transverse and traceless far away from the source, we need to project the above formula with the so-called Lambda tensor~\cite{Maggiore:2007ulw,Misner:1973prb}
\begin{eqnarray}
	\Lambda_{ij,kl} (\hat{\mathbf{k}}) = P_{ik} P_{jl} - \frac{1}{2} P_{ij} P_{kl}\,,
\end{eqnarray} 
where 
\begin{eqnarray}
	P_{ij} (\hat{\mathbf{k}}) = \delta_{ij} - \hat{k}_i \hat{k}_j\,.
\end{eqnarray}
Consequently, the general solution of tensor modes is given by
\begin{eqnarray}\label{Solhh}
	\hat{h}_{ij} (t,{\bf r}) = \frac{2(1-\gamma_W/2) {G}_N}{(1-\gamma_W) r} \Lambda_{ij,kl} (\hat{\mathbf{k}}) \frac{d^2}{dt_r^2} {\cal I}_{kl} (t_r)\,,
\end{eqnarray}
where the term proportional to the trace part $\mathcal{I}^k_{\,k}$ is dropped due to the traceless property of the Lambda tensor $\Lambda_{ij, kk} = \Lambda_{ii,kl} =0$. We have also employed Eq.~\eqref{DefGN} to write the final expression with the physically measured Newton constant $G_N$.   Moreover, it is obvious from Eq.~\eqref{Solhh} that the tensor GW solution in the GQFT is different from the GR case only by a prefactor $(1-\gamma_W/2)/(1-\gamma_W)$, which modifies the strength of the GW radiation. Unfortunately, given that the local gravity tests in the Solar system have provided strong constraints with $\gamma_W\lesssim \mathcal{O}(10^{-4})$~\cite{Gao:2024juf}, the current precision of GW detectors cannot probe such a small correction. 

\section{Gravitational Wave Detection in the GQFT}\label{SecGWDet}

In this section, we explore the detectability of different GW polarizations within the GQFT framework. To measure GW effects, we analyze the geodesic deviation equation~\cite{Carroll:2004st,Misner:1973prb,Maggiore:2007ulw}, which describes how a passing GW influences nearby free-falling particles.

For a gauge-independent measurement, we examine the relative motion of a massive particle characterized by its separation vector $S^\mu$ and four-velocity $U^\mu(x)$.  The evolution of $S^\mu$ follows: 
\begin{eqnarray}\label{EqGeoDev}
	\frac{D^2}{d\tau^2} S^\mu = R^\mu_{~\nu\rho\sigma} U^\nu U^\rho S^\sigma\,.
\end{eqnarray} 
For GWs interacting with test particles, we expand the Riemann tensor $R^\mu_{~\nu\rho\sigma}$ to first order in the metric perturbation $h_{\mu\nu}$. Assuming the particles move non-relativistically, we approximate their four-velocities as $U^\mu = (1,0,0,0)$ and proper time $\tau$ with coordinate time $t$,  {\it i.e.}, $\tau\approx t$. This simplifies the geodesic deviation equation (Eq.~\eqref{EqGeoDev}) significantly as:
\begin{eqnarray}\label{GeoDevEq}
	\frac{\partial^2 S^\mu}{\partial t^2} = R^\mu_{~00\sigma} S^\sigma\,,
\end{eqnarray}
which is totally controlled by 
\begin{eqnarray}
R_{\mu 00\sigma} = \frac{1}{2} (\partial_t^2 h_{\mu\sigma}+ \partial_\mu \partial_\sigma h_{00} - \partial_\sigma \partial_t h_{\mu 0} - \partial_t \partial_\mu h_{0\sigma} )\,. 
\end{eqnarray}
Note that, different from GR, the GQFT admits one scalar, two vector and two tensor physical GW modes. In what follows, we shall consider their respective effects on $R^\mu_{~00\sigma}$ and test particles, which mimic the influences of these polarizations on GW detectors~\cite{Maggiore:2007ulw, Carroll:2004st}. 

\subsection{Scalar Mode}

In this subsection, we focus on the single scalar mode in the GQFT. At low energies, its effects can be described by the following metric form
\begin{eqnarray}
	ds^2 = -(1+2\phi) dt^2 - 2\partial_i B dx^i dt + [(1-2\psi)\delta_{ij} + 2\partial_i \partial_j E] dx^i dx^j\,.
\end{eqnarray}
With this effective metric, we can explicitly write down components of $R^\mu_{~00\sigma}$ as follows
\begin{eqnarray}
	R_{0000} &=& R_{i000} =R_{000j} =0\,,\nonumber\\
	R_{i00j} &=& - \partial_i \partial_j \Phi - \delta_{ij} \partial_t^2 \psi+ \partial_i \partial_j \partial_t A/2 =  (\partial_i \partial_j - \delta_{ij} \partial_t^2) \psi\,,
\end{eqnarray}
where in the last equality we have used the constraints 
\begin{eqnarray}
	\Phi = -2\psi/\gamma_W\,,\quad  \partial_t A = 2(\gamma_W-2)\psi/\gamma_W\,,
\end{eqnarray}
which are shown in Ref.~\cite{Gao:2024juf} to be valid for a free-propagating scalar GW. To be more concrete, let us consider a plane wave propagating along the $z$ direction with its wave vector $k^\mu = (\omega, 0,0,\omega)$, so that the scalar wave $\psi$ can be written as
\begin{eqnarray}
	\psi = \psi_0 e^{ik_\sigma x^\sigma} = \psi_0 e^{-i\omega(t-z)}\,.
\end{eqnarray} 
Thus, the relevant components of the Riemann tensor is given by
\begin{eqnarray}
	R_{\mu00\sigma} = \left(\begin{array}{cccc}
		0 & 0& 0 & 0 \\
		0 & -\partial_t^2 \psi & 0 & 0\\
		0 & 0 & -\partial_t^2 \psi & 0\\
		0 & 0 & 0 &0
		\end{array}\right)\,.
\end{eqnarray} 
By taking this Riemann tensor into the reduced geodesic deviation equation in Eq.~\eqref{GeoDevEq}, we can solve for the evolution of the deviation vector as follows
\begin{eqnarray}\label{detectS}
	S^0 (t) &=& S^0(0) \,,\quad S^3(t) = S^3(0)\,,\nonumber\\
	S^1(t) &=& S^1(0) (1-\psi_0 e^{-i\omega(t-z)}) \,,\quad S^2(t) = S^2(0) (1-\psi_0 e^{-i\omega(t-z)}) \,,
\end{eqnarray}
which is a transverse scalar polarized mode. Note that this mode corresponds to the so-called breathing mode in the classification of GW polarizations in Refs.~\cite{Eardley:1973br,Nishizawa:2009bf}. Thus, the GW detector can be sensitive to the scalar GW, with the possible signal as shown in Eq.~\eqref{detectS}.

\subsection{Vector Mode}
As for the incoming vector GW polarization, the associated perturbed metric can be written as follows
\begin{eqnarray}
	ds^2 = -dt^2 + 2S_i  dx^i dt + 2\partial_{(i} F_{j)} dx^i dx^j\,.
\end{eqnarray}
As a result, the Riemann tensor $R^\mu_{~00\sigma}$ is given by
\begin{eqnarray}
	R_{0000} &=& R_{i000} = R_{000j} = 0\,,\nonumber\\
	R_{i00j} &=& - \partial_t \partial_{(i} [S-\partial_t F]_{j)} = 0\,, 
\end{eqnarray}
where we have used the relation $S_ i = \partial_t F_i$ for a vector GW without sources~\cite{Gao:2024juf}. Therefore, this calculation indicates that the current GW observatories cannot probe the vector GW mode in the GQFT. 

\subsection{Tensor Mode}
For the traceless-transverse tensor mode $\hat{h}_{ij}$, the Riemann tensor can be easily written down as 
\begin{eqnarray}
	R_{i00j} = \frac{1}{2} \partial_t^2 \hat{h}_{ij}\,,
\end{eqnarray}
with other components vanishing. There can be two transverse-traceless modes, labeled by $+$ and $\times$, respectively. Here we assume that a monochromatic GW propagates along the $z$ direction with a fixed frequency $\omega$. For the $+$ mode, the only nonzero components of the perturbation are $h_{11} = -h_{22} = h_+ = h^0_+ e^{-i\omega(t-z)}$ so that 
\begin{eqnarray}
	R_{\mu00\sigma} = \frac{1}{2}\left(\begin{array}{cccc}
		0 & 0& 0 & 0 \\
		0 &\partial_t^2 h_{+} & 0 & 0\\
		0 & 0 & -\partial_t^2 h_{+} & 0\\
		0 & 0 & 0 &0
	\end{array}\right)\,.
\end{eqnarray}
Thus, the solution to the geodesic deviation equation is given by
\begin{eqnarray}
	S^1(t) &=& S^1(0)(1+ \frac{1}{2}h_+^0 e^{-i\omega(t-z)})\,,\quad 	S^2(t) = S^2(0)(1- \frac{1}{2}h_+^0 e^{-i\omega(t-z)})\,, 
\end{eqnarray}
while the deviations in the $t$ and $z$ directions, $S^0$ and $S^3$, are kept invariant. On the other hand, for the $\times$ mode, the gravitational perturbations are characterized by $h_{12} = h_{21} = h_\times = h_\times^0 e^{-i\omega(t-z)}$, which leads to
\begin{eqnarray}
	R_{\mu00\sigma} = \frac{1}{2}\left(\begin{array}{cccc}
		0 & 0& 0 & 0 \\
		0 &0 & \partial_t^2 h_\times & 0\\
		0 & \partial_t^2 h_\times & 0 & 0\\
		0 & 0 & 0 &0
	\end{array}\right)\,.
\end{eqnarray} 
In this case, the geodesic deviation equation can be solved as:
\begin{eqnarray}
		S^1(t) &=& S^1(0)+ \frac{1}{2} S^2(0) h_\times^0 e^{-i\omega(t-z)}\,,\quad 	
		S^2(t) = S^2(0)+ \frac{1}{2} S^1(0) h_\times^0 e^{-i\omega(t-z)}\,, 
\end{eqnarray}
while other components of $S^\mu$ remain unchanged. 

In summary, by analyzing the geodesic deviation equations, we have demonstrated that current and near-future GW detectors, including ground-based observatories such as LIGO-Virgo-Kagra~\cite{LIGOScientific:2014pky,VIRGO:2014yos,Somiya:2011np,Aso:2013eba} and space-borne missions like LISA~\cite{LISA:2017pwj}, Taiji~\cite{Hu:2017mde}, and TianQin~\cite{TianQin:2015yph}, are capable of measuring scalar and tensor GW modes in the GQFT framework. However, due to their intrinsic properties, vector GWs remain undetectable by these laser interferometry-based detectors.

\section{Conclusion and Discussion}\label{SecConc}

GQFT provides a unified framework for gravity and the other three fundamental interactions through gauge principles. Unlike GR, this framework predicts three additional degrees of freedom in GWs: one scalar and two vector polarizations. A distinctive feature is the antisymmetric component of the EM tensor, which also contributes to GW generation.

In this work, we have investigated GW production and detection in GQFT, emphasizing the effects of these extra polarizations and novel gravitational sources. By decomposing both symmetric and antisymmetric EM tensors according to their spatial rotation properties, we demonstrate that the linearized gravitational equations separate into decoupled scalar, vector, and tensor sectors, allowing independent solutions.

Similar to GR, tensor-mode GWs are sourced by the traceless part of the matter quadrupole moment, though with a modified radiation strength. However, GQFT uniquely predicts scalar and vector gravitational radiations, both of which are absent in GR. Scalar GWs arise from two dominant sources: the trace part of the matter quadrupole moment (from the symmetric stress tensor), and an extra scalar source from the antisymmetric EM tensor, which can be generated by fermions interacting with background spin gauge field in flat spacetime approximation.

Vector GW modes, meanwhile, are exclusively sourced by the antisymmetric EM tensor, which corresponds to the fermion spin vector in the non-relativistic limit. Notably, the antisymmetric stress tensor enhances the production strengths of both scalar and vector polarizations compared to their symmetric counterparts.

To illustrate these mechanisms, we examine two astrophysical systems: a black hole binary with small orbital ellipticity (scalar GW production) and a NS binary with spin-velocity alignment (vector GW production).

Analysis of geodesic deviation equations reveals that while scalar and tensor modes can be detected by terrestrial or space-based interferometers, vector modes leave no measurable imprint on laser interferometers due to the constraint imposed by the theoretical consistency. This necessitates alternative detection strategies. Recent proposals to search for ultralight vector dark matter ~\cite{Michimura:2020vxn,Morisaki:2020gui} through measuring the oscillating forces acting on mirrors in the existing GW detectors~\cite{LIGOScientific:2021ffg,KAGRA:2024ipf,Miller:2023kkd,Frerick:2023xnf} may be adaptable to vector GW detection in GQFT. Such oscillations could arise from interactions between vector modes and fermion spins. However, a detailed exploration of this detection method lies beyond our present scope and remains an important direction for future research.

\section*{Acknowledgements}

\noindent DH is supported in part by the National Key Research and Development Program of China under Grant No. 2021YFC2203003 and No.~2024YFC2207204, by the National Astronomical Observatories, Chinese Academy of Sciences under Grant No. E4TG6601, as well as by the China Scholarship Council under Grant No. 202310740003. The work was also supported in part by the National Key Research and Development Program of China under Grant No.2020YFC2201501, the National Science Foundation of China (NSFC) under Grants No.~12147103 (special fund to the Center for Quanta-to-Cosmos Theoretical Physics), No.~11821505, and the Strategic Priority Research Program of the Chinese Academy of Sciences.

\appendix

\section{Antisymmetric Energy-Momentum Tensor for Free Fermions in the Nonrelativistic Limit}\label{uDef}

In this Appendix, we would like to derive the explicit expression for the antisymmetric EM tensor $T_{[\mu\nu]}$ for free fermions in flat spacetime limit. Note that $T_{[\mu\nu]}$ depends solely on the fermion fields in the system~\cite{Wu:2022mzr} with its expression given by
\begin{eqnarray}\label{TEMa0}
	T_{[\mu\nu]} = \frac{1}{4} \left[ \bar{\psi} \gamma_\mu i\partial_\nu \psi - \bar{\psi}\gamma_\nu i\partial_\mu \psi - i\partial_\nu \bar{\psi} \gamma_\mu \psi + i\partial_\mu \bar{\psi} \gamma_\nu \psi \right]\,,
\end{eqnarray}
which holds at the leading order in the expansion of the gravitational perturbation $h_{\mu\nu}$. This motivates us to consider the binary system of NSs, which are predominantly composed of free neutron gases. Furthermore, due to the cold and dense environment within an NS, we need only to consider neutrons in the non-relativistic limit. 

Note that the microscope world is governed by the quantum field theory and quantum mechanics. Therefore, the antisymmetric EM tensor in Eq.~\eqref{TEMa0} should be treated as an operator
$\hat{T}_{[\mu\nu]}$, while the quantity appearing in the gravitational equations is its thermal average, defined by,
\begin{eqnarray}\label{TEMaAdef}
	\langle \hat{T}_{[\mu\nu]} \rangle = \frac{1}{\cal Z}{\rm Tr}\left(\hat{\rho} \hat{T}_{[\mu\nu]}\right) \,,
\end{eqnarray}   
where $\hat{\rho}$ is the density operator for the neutron ensemble in the NS, and ${\cal Z}$ is the corresponding partition function. For our derivation, only the definition of $\langle \hat{T}_{[\mu\nu]} \rangle $ is required, the explicit forms of $\hat{\rho}$ and ${\cal Z}$ need not be specified. 
Additionally, as shown in Sec.~\ref{SecRadG}, only the $0i$ components contribute to the final expressions for gravitational radiations. Thus, in this Appendix, we focus on deriving $\langle T_{[0i]}\rangle$. 

By inserting a complete set of neutron one-particle state basis $|\mathbf{k}, r\rangle$ into Eq.~\eqref{TEMaAdef}, we can obtain
\begin{eqnarray}\label{TEMaT}
	\langle \hat{T}_{[\mu\nu]} (x) \rangle  = \frac{1}{\cal Z} \int_{\mathbf{p}}\int_{\mathbf{p}^\prime} \frac{1}{4E_{\mathbf{p}}E_{\mathbf{p}^\prime}}\sum_{s,s^\prime}  \langle \mathbf{p},s | \hat{\rho} | \mathbf{p}^\prime, s^\prime \rangle \langle \mathbf{p}^\prime, s^\prime| \hat{T}_{[\mu\nu]} | \mathbf{p},s \rangle\,,
\end{eqnarray}
where $\mathbf{p}^{(\prime)}$ and $s^{(\prime)}$ refer to the neutron momentum and spin, respectively. For the factor $\langle \mathbf{p}^\prime, s^\prime| \hat{T}_{[0i]} | \mathbf{p},s \rangle$, we can calculate it with the quantum field theory. Specifically, the quantum fields $\hat{\psi}$ and $\hat{\bar{\psi}}$ can be expanded as follows
\begin{eqnarray}\label{ExpansionPsi}
	\hat{\psi} (x) &=& \int_{\mathbf{k}} \frac{1}{\sqrt{2E_{\mathbf{k}}}} \sum_r  \left( \hat{a}^r_{\mathbf{k}} u^r (\mathbf{k}) e^{-ik\cdot x}  + \hat{b}_{\mathbf k}^{r\,\dagger} v^r(\mathbf{k}) e^{i k\cdot x}  \right) \,,\nonumber\\
	\hat{\bar{\psi}} (x) &=& \int_{\mathbf{k}} \frac{1}{\sqrt{2E_{\mathbf{k}}}} \sum_r  \left( \hat{b}^r_{\mathbf{k}} \bar{v}^r (\mathbf{k}) e^{-ik\cdot x} + \hat{a}^{r\,\dagger}_{\mathbf{k}} \bar{u}^r(\mathbf{k})e^{i k\cdot x}  \right)\,,
\end{eqnarray} 
where $\hat{a}^r_{\mathbf{k}}$ and $\hat{a}^{r\,\dagger}_{\mathbf{k}}$ are the annihilation and creation operators for neutrons of momentum $\mathbf{k}$ and spin $r$, while $\hat{b}^r_{\mathbf{k}}$ and $\hat{b}^{r\,\dagger}_{\mathbf{k}}$ are those for anti-neutrons. Therefore, by putting the expansions of Eq.~\eqref{ExpansionPsi} into the definition of $T_{[\mu\nu]}$ in Eq.~\eqref{TEMa0}, we can obtain 
\begin{eqnarray}\label{TEMa1}
	\langle \mathbf{p}^\prime, s^\prime  | \hat{T}_{[\mu\nu]} | \mathbf{p}, s\rangle = \frac{1}{2} \left[ j^{s^\prime, s}_\mu (\mathbf{p}^\prime , \mathbf{p}) P_\nu - j^{s^\prime, s}_\nu (\mathbf{p}^\prime , \mathbf{p})  P_\mu  \right] e^{iq\cdot x}\,,
\end{eqnarray}
where $P \equiv (p + p^\prime)/2$\footnote{Note that $\mathbf{P}$ here is defined as the averaged momentum of $\mathbf{p}$ and $\mathbf{p}^\prime$, which is different from its usual definition of the total momentum in the literature of dark matter direct detection~\cite{Cirelli:2013ufw}.} and $q \equiv p- p^\prime$,  while the current $j^{s^\prime, s} (\mathbf{p}^\prime, \mathbf{p})$ is defined by
\begin{eqnarray}
	j_\mu^{s^\prime, s} (\mathbf{p}^\prime, \mathbf{p}) =  \bar{u}^{s^\prime} (\mathbf{p}^\prime) \gamma_\mu u^s(\mathbf{p})\,.
\end{eqnarray}

In order to proceed and to understand its physical content in $T_{[\mu\nu]}$, we shall go to the non-relativistic limit, in which the momentum of a neutron $|\mathbf{p}|$ is much smaller than its rest mass $m$. As a result, a free neutron wave function $u^s(\mathbf{p})$ can be approximated by~\cite{Cirelli:2013ufw,Peskin:1995ev,Fitzpatrick:2012ix}
\begin{eqnarray}
	u^s({p}) = \frac{1}{\sqrt{2(E_{\mathbf{p}}+m)}}  \left( \begin{array}{c}
		(p_\mu \sigma^\mu + m)\xi^s \\
		(p_\mu \bar{\sigma}^\mu + m)\xi^s
	\end{array}	 \right) 
	\approx \frac{1}{\sqrt{4m}} \left( \begin{array}{c}
		(2m-\mathbf{p}\cdot \vec{\mathbf{\sigma}})\xi^s \\
		(2m+\mathbf{p}\cdot \vec{\mathbf{\sigma}})\xi^s
	\end{array} \right)\,,
\end{eqnarray}
where we have used the approximation $p_\mu \approx (m, \mathbf{p})$. Thus, the current $j_\mu$ is given by
\begin{eqnarray}\label{NRcurrent}
	j^{s^\prime, s}_\mu (\mathbf{p}^\prime, \mathbf{p}) = \bar{u}^{s^\prime} (p^\prime) \gamma^\mu u^s(p) \approx (2m, 2\mathbf{P}_i + 2 i (\mathbf{q}\times \mathbf{s}^{s^\prime s})_i)\,,
\end{eqnarray}
which is valid up to the first order in momenta of external particles. 
Here we have defined $\mathbf{P} = (\mathbf{p}^\prime + \mathbf{p})/2$ and $\mathbf{q} = \mathbf{p}-\mathbf{p}^\prime$ while the spin is given by
\begin{eqnarray}
	\mathbf{s}^{s^\prime s} \equiv \xi^{s^\prime \,\dagger} \frac{\vec{\mathbf{\sigma}}}{2} \xi^s \,.
\end{eqnarray}
By taking Eq.~\eqref{NRcurrent} into the matrix element of Eq.~\eqref{TEMa1}, the components of $T_{[0i]}$ in the non-relativistic limit can be simplified into
\begin{eqnarray}\label{TEMa2}
	\langle \mathbf{p}^\prime, s^\prime  | \hat{T}_{[0i]} | \mathbf{p}, s\rangle &=& \frac{1}{2} \left[j_0^{s^\prime, s} (\mathbf{p}^\prime, \mathbf{p}) \mathbf{P}_i - j_i^{s^\prime, s} (\mathbf{p}^\prime, \mathbf{p}) P_0\right] e^{i{q}\cdot {x}} \approx - i m(\mathbf{q}\times \mathbf{s}^{s^\prime s})_i e^{i\mathbf{q}\cdot \mathbf{x}}\nonumber\\
	&= & -m\epsilon_{ijk} \partial_{\mathbf{x}}^j (s^k)^{s^\prime s} e^{i\mathbf{q}\cdot \mathbf{x}}\,,
\end{eqnarray} 
where we have used the non-relativistic relations $P_\mu \approx (m, \mathbf{P})$ and $q_\mu \approx (0, \mathbf{q})$. Note that here the partial derivative $\partial_{\mathbf{x}}^j$ acts on the spatial coordinates $\mathbf{x}$.  

On the other hand, for the density matrix part, we can resort to the quantum mechanics which can provide another description of free particles. Concretely, by inserting a complete basis in the coordinate representation $\int_\mathbf{y} | \mathbf{y}\rangle \langle \mathbf{y}|  = 1$, we can transform the density matrix to the following form
\begin{eqnarray}\label{DenMat0}
	 \frac{1}{\mathcal{Z}}\langle \mathbf{p},s | \hat{\rho} | \mathbf{p}^\prime, s^\prime \rangle   &=&  \frac{1}{\mathcal{Z}} \int_\mathbf{y} \int_{\mathbf{y}^\prime} \langle  \mathbf{p}|\mathbf{y}\rangle \langle \mathbf{y},s | \hat{\rho} | \mathbf{y}^\prime, s^\prime \rangle \langle \mathbf{y}^\prime | \mathbf{p}^\prime \rangle \nonumber\\
	 &=& \frac{1}{\mathcal{Z}} \int_\mathbf{y} \int_{\mathbf{y}^\prime} \sqrt{4 E_\mathbf{p} E_{\mathbf{p}^\prime}} \langle \mathbf{y},s | \hat{\rho} | \mathbf{y}^\prime, s^\prime \rangle e^{i (\mathbf{p}^\prime \cdot\mathbf{y}^\prime - \mathbf{p}\cdot \mathbf{y})}\,,
\end{eqnarray}
where we have used the following representation transformation rules for a single particle in the quantum mechanics
\begin{eqnarray}
	\langle \mathbf{x} | \mathbf{p} \rangle = \sqrt{2E_\mathbf{p}}e^{i\mathbf{p}\cdot \mathbf{x}}\,,\quad   \langle \mathbf{p} | \mathbf{x} \rangle = \sqrt{2E_\mathbf{p}} e^{-i\mathbf{p}\cdot \mathbf{x}}\,.
\end{eqnarray}
By introducing the center-of-mass and relative coordinates
\begin{eqnarray}
 \mathbf{Y} = (\mathbf{y}+\mathbf{y}^\prime)/2\,,\quad \mathbf{z} = \mathbf{y}-\mathbf{y}^\prime\,,
\end{eqnarray}
the density matrix element in Eq.~\eqref{DenMat0} can be taken as
\begin{eqnarray}\label{DenMat1}
	\frac{1}{\mathcal{Z}}\langle \mathbf{p},s | \hat{\rho} | \mathbf{p}^\prime, s^\prime \rangle  &=& \frac{\sqrt{4 E_\mathbf{p} E_{\mathbf{p}^\prime}}}{\mathcal{Z}}\int_{\mathbf{Y}} \int_\mathbf{z} \langle \mathbf{Y}+\mathbf{z}/2,s | \hat{\rho} | \mathbf{Y}-\mathbf{z}/2, s^\prime \rangle e^{-i \mathbf{P} \cdot\mathbf{z}} e^{ -i\mathbf{q}\cdot \mathbf{Y}} \nonumber\\
	 &=& \int_\mathbf{Y} \sqrt{4 E_\mathbf{p} E_{\mathbf{p}^\prime}} W_{ss^\prime} (\mathbf{Y}, \mathbf{P}) e^{-i\mathbf{q}\cdot \mathbf{Y}}\,,
\end{eqnarray}
where $W_{ss^\prime} \equiv (1/\mathcal{Z}) \int_\mathbf{z} \langle \mathbf{Y}+\mathbf{z}/2,s | \hat{\rho} | \mathbf{Y}-\mathbf{z}/2, s^\prime \rangle e^{-i\mathbf{P}\cdot \mathbf{z}}$ is the famous Wigner function~\cite{Wigner:1932eb}, and we have also transformed integration variables into $\mathbf{Y}$ and $\mathbf{z}$. 

By combining Eqs.~\eqref{TEMa2} and \eqref{DenMat1} with Eq.~\eqref{TEMaT}, the ensemble-averaged EM tensor is obtained as,
\begin{eqnarray}
	\langle \hat{T}_{[0i]}\rangle (\mathbf{x}) &\approx& \int_\mathbf{P} \int_\mathbf{Y} \int_\mathbf{q} \sum_{s, s^\prime}  \left(-\frac{m}{\sqrt{4E_\mathbf{p}E_{\mathbf{p}^\prime}}}\right) \epsilon_{ijk} \partial^j_{\mathbf{x}} (s^k)^{s^\prime s} W_{ss^\prime} (\mathbf{Y}, \mathbf{P}) e^{i\mathbf{q}\cdot(\mathbf{x}-\mathbf{Y})} \nonumber\\
	&\approx &  -\frac{1}{2} \epsilon_{ijk} \partial^j_{\mathbf{x}}  \Xi^k(\mathbf{x})\,,
\end{eqnarray}
where we have approximated the particle energies by their masses, $E_{\mathbf{p}} \approx m$, and  defined the spin density, 
\begin{eqnarray}
	\Xi^k (\mathbf{x}) \equiv \sum_{s,s^\prime}\int_\mathbf{P} W_{ss^\prime}(\mathbf{x},\mathbf{P}) (s^k)^{s^\prime s}  \,.
\end{eqnarray}

Clearly, the ensemble averaged EM tensor components $\langle \hat{T}_{[0i]}\rangle$ is divergence free, as can be verified by the relation $\partial^i_{\mathbf{x}} \langle \hat{T}_{[0i]}\rangle (\mathbf{x}) = (-1/2) \epsilon_{ijk} \partial_{\mathbf{x}}^i \partial^j_{\mathbf{x}} \Xi^k (\mathbf{x}) =0 $. Remarkably, this result holds generally and is not restricted to the non-relativistic limit. It can also be derived directly from the general definition in Eq.~\eqref{TEMa0}:
\begin{eqnarray}
	\partial^i T_{[0i]} = \partial^\mu T_{[0\mu]} = 0\,,
\end{eqnarray}  
where the first equality follows from the antisymmetry property $T_{[00]} = 0$, and the second is obtained using the free Dirac equations $i\gamma^\mu \partial_\mu \psi = m\psi$ and $i\partial^\mu \bar{\psi} \gamma_\mu = - m\bar{\psi}$ along with $\square \psi = m^2 \psi$ and $\square \bar{\psi} = m^2 \bar{\psi}$. Furthermore, when expressing $\langle \hat{T}_{[0i]} \rangle$ in component form as in Eq.~\eqref{DefEMA}, we observe that $\Xi^k (\mathbf{x})$ contributes only to the transverse vector field $u^i$, defined as,  
\begin{eqnarray}
	u_i = -\frac{1}{2} \epsilon_{ijk} \partial^j \Xi^k\,,
\end{eqnarray}
while the longitudinal part, parametrized by the scalar $s$, vanishes. However, this conclusion applies strictly to free fermions. In the presence of additional gauge interactions among fermions, $s$ can generally become nonzero, for that we will explicitly demonstrate in the next Appendix.
 
Finally, although we have focused on deriving $\langle T_{[0i]} \rangle$, we would like to emphasize that the method presented here is general and can be extended to compute other ensemble averaged components of the EM tensor.

\section{Longitudinal Component of $T_{[0i]}$ for Interacting Fermions}\label{SDef}

In Appendix \ref{uDef}, we demonstrated that free fermions can generate non-trivial contributions to the vector component $u_i$ in the antisymmetric EM tensor $T_{[0i]}$ defined in Eq.~\eqref{DefEMA}. However, a limitation of this derivation is that the scalar component $s$ vanishes identically in this scenario.

As highlighted in Refs.~\cite{Wu:2022aet,Wu:2022mzr}, the complete antisymmetric EM tensor for fermions interacting with background gauge fields should instead be defined as:
\begin{eqnarray}
	T_{[\mu\nu]} = \frac{1}{4} \left[ \bar{\psi} \gamma_\mu iD_\nu \psi -\bar{\psi}\gamma_\nu i D_\mu \psi - i D_\nu \bar{\psi} \gamma_\mu \psi + i D_\mu \bar{\psi}\gamma_\nu \psi \right]\,, 
\end{eqnarray} 
where $D_\mu \psi \equiv (\partial_\mu - i g \mathcal{A}_\mu) \psi $ and $D_\mu \bar{\psi} \equiv \partial_\mu \bar{\psi} + ig \bar{\psi} \mathcal{A}_\mu $ are the covariant derivatives for the fermion $\psi$ and $\bar{\psi}$, respectively.  Here, ${\cal A}_\mu = \mathcal{A}^a_\mu t^a$ represents the gauge field, with $t^a$ denoting the Hermitian generators of the gauge group. 

This raises an important question: Can a non-trivial gauge field background $\mathcal{A}^a_\mu$ induce a non-zero longitudinal component $T_{[0i]}$, thereby sourcing scalar gravitational wave (GW) radiation? We address this question in the present appendix.

Before proceeding, we note that the fermion Dirac equation in the presence of $\mathcal{A}^a_\mu$ takes the form,
\begin{eqnarray}
	(i\gamma^\mu D_\mu -m) \psi &=& i\slashed{\partial}\psi + (g\slashed{\mathcal{A}}^a t^a - m) \psi = 0\,,\nonumber\\
	-i D_\mu \bar{\psi}\gamma^\mu  - m \bar{\psi} &=& -i\partial_\mu \bar{\psi} \gamma^\mu + \bar{\psi} (g \slashed{\mathcal{A}}^a t^a -m) = 0\,,
\end{eqnarray}
which leads to the following identities:
\begin{eqnarray}\label{identity2}
	(i\slashed{D})(i\slashed{D}) \psi &=& m^2 \psi = - D^\mu D_\mu \psi + \frac{g}{2} \sigma^{\mu\nu} \mathcal{F}_{\mu\nu} \psi\,, \nonumber\\
	D^\nu D^\mu \bar{\psi} (-i\gamma^\mu) (-i\gamma^\nu) & = & m^2\bar{\psi} = -D^\mu D_\mu \bar{\psi} + \frac{g}{2}  \bar{\psi}\sigma^{\mu\nu}\mathcal{F}_{\mu\nu}\,,
\end{eqnarray}
where we define,
\begin{eqnarray}
	\sigma^{\mu\nu}  = \frac{i}{2}[\gamma^\mu, \gamma^\nu]\,, \quad \mathcal{F}_{\mu\nu} = \partial_\mu \mathcal{A}_\nu - \partial_\nu \mathcal{A}_\mu -ig \left[\mathcal{A}_\mu,\mathcal{A}_\mu\right]\,. 
\end{eqnarray}

Now let us consider the spatial divergence of $T_{[0i]}$. Owing to the antisymmetry of subscripts of $T_{[0i]}$, there is not the component $T_{[00]}$, so that we can write 
\begin{eqnarray}
    &&\partial^i T_{[0i]} = \partial^\mu T_{[0\mu]} \nonumber\\
    &=& \frac{1}{4} \partial^\mu \left[ \bar{\psi} \gamma_0 iD_\mu \psi -\bar{\psi}\gamma_\mu iD_0\psi - iD_\mu \bar{\psi} \gamma_0 \psi + iD_0 \bar{\psi} \gamma_\mu \psi \right]\,.
\end{eqnarray}
Since every term in the second line is gauge invariant, we can represent them solely in terms of the covariant derivatives. Let us take the first one $\partial^\mu (\bar{\psi}\gamma_0 iD_\mu \psi)$ as an example:
\begin{eqnarray}
	&& \partial^\mu (\bar{\psi}\gamma_0 iD_\mu\psi) = \partial^\mu \bar{\psi} \gamma_0 iD_\mu \psi + \bar{\psi}\gamma_0 \partial^\mu (iD_\mu\psi) \nonumber\\
	&=& \partial^\mu \bar{\psi}\gamma_0 iD_\mu \psi + \bar{\psi} (ig\mathcal{A}_\mu) \gamma_0 iD_\mu \psi + \bar{\psi}\gamma_0 \partial^\mu (iD_\mu\psi) -\bar{\psi} \gamma_0 (ig\mathcal{A}_\mu) iD_\mu \psi \nonumber\\
	&=& D^\mu \bar{\psi}\gamma_0 iD_\mu \psi + \bar{\psi}\gamma_0 D^\mu (iD_\mu \psi)\,,
\end{eqnarray}
where we have added and subtracted exactly the same term $\bar{\psi} (ig\mathcal{A}_\mu) \gamma_0 iD_\mu \psi$. In effect, we have replaced the partial derivative with the corresponding gauge covariant derivative in this formula. Following the same procedure for other three terms, we can write $\partial^i T_{[0i]}$ as  
\begin{eqnarray}
	&& \partial^i T_{[0i]} = \partial^\mu T_{[0\mu]}  \\
	&=& \frac{1}{4}\left[ ( \bar{\psi}\gamma_0 i D^2\psi - iD^2 \bar{\psi}\gamma_0 \psi )- (iD^\mu\bar{\psi}\gamma_\mu D_0 \psi +\bar{\psi}\slashed{D}iD_0\psi) + (i D_\mu D_0 \bar{\psi} \gamma^\mu \psi + i D_0\bar{\psi}\slashed{D}\psi ) \right]\,,\nonumber
\end{eqnarray}
where we have regrouped the contributions into three terms as shown in different parenthesis.

We shall compute them one by one.
\begin{itemize}
	\item Term 1:
	\begin{eqnarray}
		\bar{\psi}\gamma_0 iD^2 \psi - i D^2 \bar{\psi} \gamma_0 \psi = \frac{ig}{2} \left( \bar{\psi}\gamma_0 \sigma^{\mu\nu}t^a\psi -\bar{\psi} \sigma^{\mu\nu}\gamma_0 t^a \psi  \right)\mathcal{F}^a_{\mu\nu}\,,
	\end{eqnarray}
	where we have used the relations in Eq.~\eqref{identity2} to simplify the expression. By employing the following identity
	\begin{eqnarray}
		\gamma_0 \sigma^{\mu\nu} = 2i (\delta^\mu_0 \gamma^\nu - \delta^\nu_0 \gamma^\mu) + \sigma^{\mu\nu}\gamma_0\,.
	\end{eqnarray} 
	this formula can be further reduced to
	\begin{eqnarray}
		&& \bar{\psi}\gamma_0 iD^2 \psi - i D^2 \bar{\psi} \gamma_0 \psi  = (ig/2) \times2i \left( \delta^\mu_0 \bar{\psi}\gamma^\nu t^a\psi - \delta^\nu_0 \bar{\psi}\gamma^\mu t^a \psi \right) \mathcal{F}^a_{\mu\nu} \nonumber\\
		&=& -2g \bar{\psi}\gamma^i t^a\psi \mathcal{F}^a_{0i}\,.
	\end{eqnarray}
	\item Term 2:
	\begin{eqnarray}
		&& -(iD^\mu \bar{\psi} \gamma_\mu D_0 \psi + \bar{\psi} \slashed{D}i D_0 \psi ) = m\bar{\psi} D_0 \psi - \bar{\psi} i\gamma^\mu [D_\mu, D_0]\psi - \bar{\psi} D_0 i\slashed{D}\psi \nonumber\\
		& =& g \bar{\psi} \gamma^i t^a\psi \mathcal{F}^a_{0i}\,.
	\end{eqnarray}
	\item Term 3:
	\begin{eqnarray}
		&& iD_\mu D_0 \bar{\psi}\gamma^\mu \psi + iD_0 \bar{\psi}\slashed{D}\psi = i [D_\mu, D_0] \bar{\psi}\gamma^\mu \psi + iD_0 D_\mu \bar{\psi}\gamma^\mu \psi + iD_0 \bar{\psi}\slashed{D}\psi \nonumber\\
		&=& i\bar{\psi}\gamma^\mu t^a \psi (ig\mathcal{F}^a_{\mu 0}) = g\bar{\psi} \gamma^i t^a \psi \mathcal{F}^a_{0i}\,.
	\end{eqnarray}
\end{itemize} 
Upon summing these three terms, we observe that all contributions cancel exactly. Consequently, the spatial divergence of $T_{[0i]}$ vanishes identically, i.e., $\partial^i T_{[0i]}=0$. From its definition in Eq.~\eqref{DefEMA}, the longitudinal component $s$ relates to $T_{[0i]}$ through 
\begin{eqnarray} \label{sLap0}
	  \partial^i \partial_i s = -\partial^i T_{[0i]}=0  \,,
\end{eqnarray}
which demonstrates that $s$ must vanish identically for charged fermions, even in the presence of conventional gauge field backgrounds. This result implies that the scalar GW source $\widetilde{S}$ cannot be generated in this case, as evidenced by the relation in Eq.~\eqref{IntS}.

While this derivation holds for conventional gauge theories in quantum field theory, the situation differs for spin gauge field in GQFT. When including the background spin gauge field $A_{\mu a b}\Sigma^{ab}/2$, where $\Sigma^{ab} = (i/4)[\gamma^a, \gamma^b]$ represents the group generators of spin gauge symmetry SP(1,3) in the spinor representation, the covariant derivative takes the form:
\begin{eqnarray}
	D_\mu \psi = \partial_\mu \psi- \frac{i}{2} g_G A_{\mu ab}\Sigma^{ab} \psi\,,
\end{eqnarray}
where $g_G$ is the corresponding gauge coupling constant. 
As a result,  the commutator of the associated covariant derivatives gives~\cite{Wu:2022mzr}
\begin{eqnarray}
	i[D_\mu, D_\nu] = g_G F_{\mu\nu}^{~~ab}\Sigma_{ab} /2\,,
\end{eqnarray}
where the field strength is defined by,
\begin{eqnarray}
	F_{\mu\nu}^{~~ab} \equiv \partial_\mu A_\nu^{~ab} - \partial_\nu A_\mu^{~ab} + g_G A_{\mu~c}^{~a} A_{\nu}^{~cb} - g_G A_{\nu~c}^{~a} A_{\mu}^{~cb}\,.
\end{eqnarray}

In flat spacetime, we treat spatial coordinate indices (denoted by Greek letters $\mu$, $\nu$, ...) and internal spin vector indices (denoted by Latin letters $a$, $b$,...) equivalently. However, for clarity of presentation, we maintain distinct notation to track their origins. The fermion equations of motion become:
\begin{eqnarray}
	i\gamma^\mu D_\mu \psi \equiv i \gamma^\mu \partial_\mu \psi + \frac{g_G}{2} \gamma^\mu A_{\mu ab} \Sigma^{ab} \psi = m \psi\,, \nonumber\\
	-i D_\mu \bar{\psi} \gamma^\mu \equiv -i\partial_\mu \bar{\psi} \gamma^\mu + \frac{g_G}{2} A_{\mu ab} \bar{\psi} \Sigma^{ab} \gamma^\mu = m \bar{\psi}\,.
\end{eqnarray} 
We emphasize that the commutators $[\gamma^\mu, \Sigma^{ab}]$ are non-vanishing because $\Sigma^{ab}$ itself is composed of gamma matrices. This leads to the following important identities:
\begin{eqnarray}
	m^2 \psi &=& (i\gamma^\nu D_\nu) (i\gamma^\mu D_\mu) \psi = \gamma^\nu \gamma^\mu (iD_\nu) (iD_\mu) \psi + \frac{i}{2} g_G A_{\nu cd} \gamma^\nu [\Sigma^{cd}, \gamma^\mu] D_\mu \psi \nonumber\\
	&=& (iD_\mu)(iD^\mu)\psi + \frac{1}{2}g_G  F_{\mu\nu}^{~~ab} \Sigma^{\mu\nu} \Sigma_{ab} \psi - \frac{i}{2} g_G A_{\nu ab} \gamma^\nu [\gamma^\mu, \Sigma^{ab}] D_\mu \psi\,,\nonumber\\
	m^2 \bar{\psi} &=& (-i D_\nu) (-iD_\mu\bar{\psi}\gamma^\mu)\gamma^\nu = (-iD_\nu) (-iD_\mu) \bar{\psi} \gamma^\mu \gamma^\nu - \frac{i}{2} g_G D_\mu \bar{\psi} [\gamma^\mu, \Sigma^{ab}]  A_{\nu ab}\gamma^\nu \nonumber\\
	&=& (-iD_\mu)(-iD^\mu)\bar{\psi} +\frac{1}{2} g_G F_{\mu\nu}^{~~ab} \bar{\psi} \Sigma_{ab} \Sigma^{\mu\nu} -\frac{i}{2} g_G D_\mu \bar{\psi} [\gamma^\mu, \Sigma^{ab}] A_{\nu ab}\gamma^\nu\,.
\end{eqnarray}

We now turn to the calculation of the divergence of $T_{[0i]}$, which can be transformed into
\begin{eqnarray}
	\partial^i T_{[0i]} = \partial^\mu T_{[0\mu]} = \frac{1}{4} \partial^\mu \left[ \bar{\psi} \gamma_0 i D_\mu \psi - \bar{\psi} \gamma_\mu i D_0 \psi - i D_\mu \bar{\psi} \gamma_0 \psi + i D_0 \bar{\psi} \gamma_\mu \psi \right]\,. 
\end{eqnarray}
Here there are four terms in this expression, and we shall compute them one by one
\begin{itemize}
	\item Term 1: 
	\begin{eqnarray}
		&&\partial^\mu \left( \bar{\psi}\gamma_0 i D_\mu \psi \right) = \partial^\mu \bar{\psi} \gamma_0 i D_\mu \psi + \bar{\psi} \gamma_0 i \partial^\mu D_\mu \psi \nonumber\\
		&=& D^\mu \bar{\psi} \gamma_0 i D_\mu \psi + \bar{\psi}\gamma_0 i D^\mu D_\mu \psi + \frac{i}{2} g_G A_{\mu ab} \bar{\psi} [\gamma_0 ,\Sigma^{ab} ] i D^\mu \psi  \nonumber\\
		&=& i D^\mu \bar{\psi} \gamma_0 D_\mu \psi -i m^2 \bar{\psi} \gamma_0 \psi + \frac{i}{2} g_G F_{\mu\nu}^{~~ab}  \bar{\psi} \gamma_0 \Sigma^{\mu\nu}\Sigma_{ab} \psi \nonumber\\
		&& + \frac{1}{2} g_G A_{\nu ab} \bar{\psi} \gamma_0 \gamma^\nu [\gamma^\mu, \Sigma^{ab}] D_\mu \psi  - \frac{1}{2} g_G A_{\mu ab} \bar{\psi} [\gamma_0, \Sigma^{ab}] D^\mu \psi\,.
	\end{eqnarray}
	
	\item Term 2:
	\begin{eqnarray}
		&&\partial^\mu \left( \bar{\psi} \gamma_\mu i D_0 \psi \right) = \partial^\mu \bar{\psi} \gamma_\mu i D_0 \psi + \bar{\psi} \gamma_\mu i \partial^\mu D_0 \psi \nonumber\\
		&=& D^\mu \bar{\psi} \gamma_\mu iD_0 \psi + \bar{\psi} \gamma_\mu i D^\mu D_0 \psi + \frac{i}{2}g_G A_{\mu ab} \bar{\psi} [\gamma^\mu,\Sigma^{ab}] i D_0 \psi \nonumber\\
		&=& - m \bar{\psi} D_0 \psi + \bar{\psi} \gamma^\mu i [D_\mu, D_0] \psi + \bar{\psi} \gamma^\mu D_0 i D_\mu \psi - \frac{1}{2} g_G A_{\mu ab} \bar{\psi} [\gamma^\mu, \Sigma^{ab}] D_0 \psi \nonumber\\
		&=& -\frac{1}{2} g_G F_{0\mu}^{~~ab} \bar{\psi} \gamma^\mu \Sigma_{ab} \psi + \frac{1}{2}g_G A_{0ab} \bar{\psi}[\gamma^\mu, \Sigma^{ab}]  D_\mu \psi  - \frac{1}{2} g_G A_{\mu ab} \bar{\psi} [\gamma^\mu,\Sigma^{ab}] D_0\psi\,.
	\end{eqnarray}
	
	\item Term 3:
	\begin{eqnarray}
		&& \partial^\mu \left( i D_\mu \bar{\psi}\gamma_0 \psi\right) = i\partial^\mu D_\mu \bar{\psi} \gamma_0 \psi + i D_\mu \bar{\psi} \gamma_0 \partial^\mu \psi \nonumber\\
		&=& iD^\mu D_\mu \bar{\psi}\gamma_0\psi + iD_\mu \bar{\psi} \gamma_0 D^\mu \psi + \frac{1}{2} g_G A_{\mu ab} D^\mu \bar{\psi} [\Sigma^{ab}, \gamma_0] \psi \nonumber\\
		&=& i D_\mu \bar{\psi} \gamma_0 D^\mu \psi -im^2 \bar{\psi}\gamma_0 \psi +\frac{i}{2} g_G F_{\mu\nu}^{~~ab} \bar{\psi}\Sigma_{ab} \Sigma^{\mu\nu}\gamma_0 \psi \nonumber\\
		&& + \frac{1}{2}g_G A_{\nu ab} D_\mu\bar{\psi} [\gamma^\mu,\Sigma^{ab}]\gamma^\nu \gamma_0 \psi -\frac{1}{2} g_G A_{\mu ab} D^\mu \bar{\psi} [\gamma_0, \Sigma^{ab}] \psi\,.
	\end{eqnarray}
	
	\item Term 4:
	\begin{eqnarray}
		&& \partial^\mu \left( iD_0 \bar{\psi} \gamma_\mu \psi \right) = i \partial^\mu D_0 \bar{\psi} \gamma_\mu \psi + iD_0 \bar{\psi} \gamma_\mu \partial^\mu \psi \nonumber\\
		&=& iD^\mu D_0 \bar{\psi} \gamma_\mu \psi + iD_0 \bar{\psi} \gamma_\mu D^\mu \psi + \frac{1}{2} g_G A_{\mu ab} D_0 \bar{\psi}[\Sigma^{ab},\gamma^\mu]\psi \nonumber\\
		&=& i [D_\mu, D_0] \bar{\psi} \gamma^\mu \psi + (i D_0 D_\mu \bar{\psi})\gamma^\mu \psi + m D_0\bar{\psi} \psi - \frac{1}{2} g_G A_{\mu ab} D_0 \bar{\psi} [\gamma^\mu, \Sigma^{ab}] \psi \nonumber\\
		&=& \frac{1}{2} g_G F_{0\mu}^{~~ab} \bar{\psi} \Sigma_{ab} \gamma^\mu \psi  + \frac{1}{2} g_G A_{0ab} D_\mu \bar{\psi} [\gamma^\mu, \Sigma^{ab}] \psi -\frac{1}{2} g_G A_{\mu ab} D_0 \bar{\psi} [\gamma^\mu, \Sigma^{ab}] \psi\,.
	\end{eqnarray}
\end{itemize}
By putting the above four terms together, we can obtain
\begin{eqnarray}\label{pT0i}
	\partial^i T_{[0i]} &=& \frac{g_G}{4}\left(\frac{i}{4} F_{\mu\nu}^{~~ab} \bar{\psi} \left[ \{\gamma_0, \Sigma^{\mu\nu}\}, \Sigma_{ab} \right]\psi  + m A_{\mu ab} \bar{\psi}\{\Sigma_0^{~\mu},\Sigma^{ab}\} \psi  \right. \nonumber\\
	&& + i A_{\nu ab} \left( \bar{\psi} \Sigma^{\mu\nu} \gamma_0 \Sigma^{ab} D_\mu \psi - D_\mu \bar{\psi} \Sigma^{ab} \gamma_0 \Sigma^{\mu\nu} \psi \right) \nonumber\\
	&& \left. + \frac{1}{2} A_{\mu ab} \left( \bar{\psi} \Sigma^{ab} \gamma_0 D^\mu \psi  + D^\mu \bar{\psi} \gamma_0 \Sigma^{ab} \psi - \bar{\psi} \Sigma^{ab} \gamma^\mu D_0 \psi - D_0 \bar{\psi} \gamma^\mu \Sigma^{ab} \psi \right)  \right)\,,
\end{eqnarray}
where we have used the fermion equations of motion and the gamma matrix algebra to simplify the expression. On the other hand, according to  Eq.~\eqref{DefEMA}, $\partial^i T_{[0i]}$ only picks up the longitudinal component $s$
\begin{eqnarray} \label{sLap}
	\partial^i T_{[0i]} = - \partial^i \partial_i s\,,
\end{eqnarray}
where the part related to $u_i$ vanishes by using its definition $\partial^i u_i = 0$. Note that Eq.~\eqref{sLap} can be viewed as the Poisson equation for $s$. As a result, $s$ can be solved as follows
\begin{eqnarray}
	s (\mathbf{r}) = \frac{1}{4\pi} \int_{\mathcal{V}} d^3 y \frac{\partial^i T_{[0i]} (\mathbf{y})}{|\mathbf{r} - \mathbf{y}|}\,,
\end{eqnarray} 
where $\partial^i T_{[0i]}$ can take its concrete expression in Eq.~\eqref{pT0i}. From this derivation, it is seen that the interaction of fermions with background spin gauge fields can indeed produce nontrivial contribution to $s$ in $T_{[0i]}$. 

Given the general solutions to the scalar GW production in Eqs.~\eqref{SolPsi} and \eqref{SolPhi}, the longitudinal source $s$ can generate scalar GWs via the following quantity:
\begin{eqnarray}\label{Ssol}
	\widetilde{S} = 3 \int_\mathcal{V} d^3y s (\mathbf{y}) = \int_\mathcal{V} d^3 y y^i T_{[0i]} = -\frac{1}{2} \int_{\mathcal{V}} d^3y y^i y_i \partial^j T_{[0j]}\,,
\end{eqnarray} 
where we have used the following identity
\begin{eqnarray}
	&&\int_\mathcal{V} d^3 y y^j T_{[0j]} = \int_\mathcal{V} d^3 y y^i \delta^j_i T_{[0j]} = \int_\mathcal{V} d^3y y^i (\partial^j y_i) T_{[0j]} \nonumber\\
	&=& -\int_{\mathcal{V}} d^3y  y^j T_{[0j]} - \int_\mathcal{V} d^3y y^i y_i \partial^j T_{[0j]} \nonumber\\
	&=& - \frac{1}{2} \int_\mathcal{V} d^3y y^i y_i \partial^j T_{[0j]}\,,
\end{eqnarray}
where the equality in the second line comes from the integration by parts with getting rid of terms of total divergence. Eq.~\eqref{Ssol} indicates that, for a given background spin gauge field $A_{\mu ab}$ and its associated field strength ${F}_{\mu\nu}^{~~ab}$,  we can directly compute the source $\widetilde{S}$ for the scalar GW mode production without resorting to its local origin $s$. We have commented its importance in the generation of scalar GWs in Sec.~\ref{SecRadS}.

\section{Proof of Several Useful Relations}\label{Proofs}
Here we would like to use the long-wavelength expansion technique~\cite{Will:1977wq,Misner:1973prb,Will:1993hxu} to prove several useful relations of Eqs.~\eqref{ScalarCons} and \eqref{ConsistencyV} in the main text. 
\subsection{Relations for Scalar Fields}\label{ProofScalar}
We shall firstly focus on the scalar field relation in Eq.~\eqref{ScalarCons}
\begin{eqnarray}
    	M = \int_{\cal V} d^3 y \left[\rho + \partial_{\tilde{t}_r} (b-s)\right](\tilde{t}_r, \mathbf{y})\,,
\end{eqnarray}
which is crucial to show the consistency in the derivation of the scalar GW radiation formulae. Note that all fields in the volume integrand are functions of retarded time $\tilde{t}_r = t-|\mathbf{r}-\mathbf{y}| \simeq t_r + \hat{\mathbf{k}}\cdot \mathbf{y}$, which gives extra $\mathbf{y}$ dependence and should be taken care of when performing the volume integration. Here we have already kept the leading-order term in the far-field expansion~\cite{Will:1977wq} in $\tilde{t}_r$, where $t_r = t-r$ and $\hat{\mathbf{k}} = \hat{\mathbf{r}}$ denotes the unit vector along the propagation of GWs of wavenumber $\mathbf{k}$.

When the scalar GW wavelength $\lambda \sim 1/\partial_t $ is much larger than the typical source length scale, we can regard $(\hat{\mathbf{k}}\cdot \mathbf{y}) \partial_{t_r}$ as a small expanding parameter. Hence, we can perform this long-wavelength expansion for the matter field $\rho(\tilde{t}_r, \mathbf{y})$  as follows
\begin{eqnarray}
	\rho (\tilde{t}_r, \mathbf{y}) = \sum_{n=0}^\infty \frac{1}{n!} \left( \hat{\mathbf{k}}\cdot \mathbf{y}\right)^n  \frac{\partial^n}{\partial t_r^n} \rho(t_r,\mathbf{y})\,.
\end{eqnarray}
In each term of this expansion, the time variable of fields $\rho(t_r, \mathbf{y})$ turns to $t_r$, which does not depend on the integration variable $\mathbf{y}$ any more. We can perform the volume integration for the terms in this series:
\begin{itemize}
	\item $n=0$: 
	\begin{eqnarray}
		\int_{\cal V} d^3 y \rho(t_r, \mathbf{y}) = M\,.
	\end{eqnarray}
	\item $n=1$:
	\begin{eqnarray}
		\int_{\cal V} d^3 y \,\hat{{k}}_i y^i \partial_{t_r} \rho (t_r, \mathbf{y})= -\hat{k}_i \int_{\cal V} d^3y y^i \partial_j \partial^j (b-s)  = \hat{k}_i \int_{\cal V} d^3 y \partial^i (b-s) =0\,,
	\end{eqnarray}
	where we have used the EM conservation in Eq.~\eqref{EqC0} in the first equality and applied the integration by parts in the second one. 
	\item $n\geqslant 2 $ :
	\begin{eqnarray}
		&&\sum_{n=2}^\infty \frac{1}{n!} \, \hat{k}_{i_1} \hat{k}_{i_2} \cdots \hat{k}_{i_n} \int_{\cal V} d^3 y y^{i_1} y^{i_2} \dots y^{i_n} \partial_{t_r}^n \rho(t_r, \mathbf{y})\nonumber\\
		&=& -\sum_{n=2}^\infty \frac{1}{n!} \, \hat{k}_{i_1} \hat{k}_{i_2} \cdots \hat{k}_{i_n} \int_{\cal V} d^3 y y^{i_1} y^{i_2} \dots y^{i_n} \partial_{t_r}^{n-1} \left[\partial_j \partial^j (b-s)\right] \nonumber\\
		&=& \sum_{n=2}^\infty \frac{1}{(n-2)!} \hat{k}_{i_1} \hat{k}_{i_2} \cdots \hat{k}_{i_{n-2}} \int_{\cal V} d^3 y y^{i_1} y^{i_2} \cdots y^{i_{(n-2)}} \partial_{t_r}^{n-2} [\partial_{t_r} (b-s)] \nonumber\\
		&=& - \int_{\cal V} d^3 y \partial_{\tilde{t}_r} (b-s) (\tilde{t}_r, \mathbf{y})\,,
	\end{eqnarray}
\end{itemize}
By combining the above results, we can obtain 
\begin{eqnarray}
	\int_{\cal V} d^3 y \rho(\tilde{t}_r,\mathbf{y}) = M - \int_{\cal V} d^3 y \partial_{\tilde{t}_r} (b-s) (\tilde{t}_r, \mathbf{y})\,,
\end{eqnarray}
which is exactly the formula in Eq.~\eqref{ScalarCons} we are pursuing. 

\subsection{Relations for Vector Fields}\label{ProofVector}
We would like to provide explicit proofs for two consistency relations in Eq.~\eqref{ConsistencyV}, which are reproduced here
\begin{eqnarray}\label{ConV1}
	\int_{\cal V} d^3 y  (v_i - \partial_{\tilde{t}_r} \theta_i) (\tilde{t}_r,\mathbf{y}) =0\,,
\end{eqnarray}
and
\begin{eqnarray}\label{ConV2}
	\hat{k}_j\int_{\cal V} d^3 y y^j (v_i -\partial_t \theta_i) \simeq -\frac{1}{2} \epsilon_{ijk} \hat{k}^j L^k\,.
\end{eqnarray}
Note that the time variable of the integrand in Eq.~\eqref{ConV1} is the retarded time $\tilde{t}_r = t- |\mathbf{x} -\mathbf{y}|$ which introduces an extra $\mathbf{y}$ dependence. On the other hand, the notation $\simeq$ means that Eq.~\eqref{ConV2} is only valid in the long-wavelength limit, {\it i.e.}, it is true up to the leading order in the small $(\hat{\mathbf{k}}\cdot\mathbf{y})\partial_{t_r}$ expansion. Before delving into the proofs of the above two equations, we would like to mention one new constraint on the vector matter sources 
\begin{eqnarray}\label{ConstVe}
	\partial_j \partial^j u_i + \partial_t \left( \epsilon_{ijk} \partial^j \omega^k\right) =0\,,
\end{eqnarray}
which can be derived by combining Eqs.~\eqref{EqH0iV} and \eqref{EqHij}. 

In order to prove Eq.~\eqref{ConV1}, one just needs to compute the following volume integral
\begin{eqnarray}
	\int_{\mathcal{V}} d^3y v_i(\tilde{t}_r, \mathbf{y}) = \sum_{n=0}^\infty \frac{1}{n!} \hat{k}_{i_1} \cdots \hat{k}_{i_n} \int_\mathcal{V}d^3y  y^{i_1} \cdots y^{i_n} \partial_{t_r}^{n} v_i (t_r, \mathbf{y})\,.
\end{eqnarray}
For the above summation, we need to deal with the term of $n=0$ and other terms separately:
\begin{itemize}
	\item $n=0$: 
	\begin{eqnarray}\label{PConVt}
		\int_\mathcal{V} d^3 y v_i(t_r, \mathbf{y}) =0\,.
	\end{eqnarray}
	\item $n\geqslant 1$:
	\begin{eqnarray}\label{PConV0}
		&&\sum_{n=1}^\infty \frac{1}{n!} \hat{k}_{i_1} \cdots \hat{k}_{i_n} \int_\mathcal{V}d^3y  y^{i_1} \cdots y^{i_n} \partial_{t_r}^{n-1} \Big(\partial_{t_r} v_i (t_r, \mathbf{y})\Big) \nonumber\\
		&=& \sum_{n=1}^\infty \frac{1}{n!} \hat{k}_{i_1} \cdots \hat{k}_{i_n} \int_\mathcal{V}d^3y  y^{i_1} \cdots y^{i_n}  \partial^{n-1}_{t_r} \left(\partial_j \partial^j \theta_i -\epsilon_{ijk} \partial^j \omega^k -\partial_{t_r} u_i \right)\,,
	\end{eqnarray}
	where we have used the matter conservation equation of Eq.~\eqref{EqCiV}. 
	In what follows, we shall perform the integration for terms in the parenthesis one by one. 
	\begin{eqnarray}\label{PConV1}
		&& \sum_{n=1}^\infty \frac{1}{n!} \hat{k}_{i_1} \cdots \hat{k}_{i_n} \int_\mathcal{V}d^3y  y^{i_1} \cdots y^{i_n}  \partial^{n-1}_{t_r}  \partial_j \partial^j \theta_i  \nonumber\\
		&=& \sum_{n=2}^\infty \frac{1}{(n-2)!} \hat{k}_{i_1} \cdots \hat{k}_{i_{(n-2)}} \int_\mathcal{V} d^3 y y^{i_1} \cdots  y^{i_{(n-2)}} \partial_{t_r}^{n-1} \theta_i = \int_\mathcal{V}d^3y  \partial_{\tilde{t}_r}\theta_i (\tilde{t}_r, \mathbf{y})\,,
	\end{eqnarray} 
	where we have applied the integration by parts in the first equality and used the definition of the long-wavelength expansion in the second. 
	For the integral of the term involving $\omega^k$, we have
	\begin{eqnarray}\label{PConV2}
		&& -\sum_{n=1}^\infty \frac{1}{n!} \hat{k}_{i_1} \cdots \hat{k}_{i_n} \int_\mathcal{V} d^3 y y^{i_1} \cdots y^{i_n} \partial_{t_r}^{n-1} \left( \epsilon_{ijk} \partial^j \omega^k \right)  \nonumber\\
		&=& \sum_{n=1}^\infty \frac{1}{n!} \hat{k}_{i_1} \cdots \hat{k}_{i_n} \int_\mathcal{V} d^3 y y^{i_1} \cdots y^{i_n} \partial_{t_r}^{n-2}\left( \partial^j \partial_j u_i\right)\nonumber\\
		&=& \sum_{n=2}^\infty \frac{1}{(n-2)!} \hat{k}_{i_1} \cdots \hat{k}_{i_{(n-2)}} \int_\mathcal{V} d^3 y y^{i_1} \cdots y^{i_(n-2)} \partial_{t_r}^{n-2} u_i 
		 = \int_\mathcal{V} d^3 y u_i (\tilde{t}_r, \mathbf{y})\,,
	\end{eqnarray}
	where we have applied Eq.~\eqref{ConstVe}  in the first equality. 
	Finally, according to the definition of long-wavelength expansion, the last term related to $u_i$ directly gives 
	\begin{eqnarray}
		\sum_{n=1}^\infty \frac{1}{n!} \hat{k}_{i_1} \cdots \hat{k}_{i_n} \int_\mathcal{V} d^3 y y^{i_1} \cdots y^{i_n} \partial^n_{t_r} u_i = \int_\mathcal{V} d^3 y u_i (\tilde{t}_r, \mathbf{y})\,,
	\end{eqnarray}
	in which we have added a vanishing term with $n=0$. 
	Therefore, by taking Eqs.~\eqref{PConV1} and \eqref{PConV2} into Eq.~\eqref{PConV0} and considering Eq.~\eqref{PConVt}, we can yield
	\begin{eqnarray}
		\int_\mathcal{V} d^3y v_i(\tilde{t}_r, \mathbf{y})  = \int_\mathcal{V} d^3y \partial_{\tilde{t}_r} \theta_i(\tilde{t}_r, \mathbf{y})\,,
	\end{eqnarray}
	which is nothing but just the wanted relation in Eq.~\eqref{ConV1}. 
\end{itemize}

As for the relation in Eq.~\eqref{ConV2}, we shall still apply the long-wavelength expansion technique. But  either $\epsilon_{ijk} \hat{k}^j L^k$ or $\hat{k}^j \int_\mathcal{V} d^3 y y_j (v_i-\partial_t \theta_i)$ appears as coefficients of terms of $\mathcal{O}(1/r^2)$ in the far-field expansion as evident in Eqs.~\eqref{EqHGV1} and \eqref{EqHGV2}. Therefore, we should keep the time-independent components, which correspond to the leading-order terms in the long-wavelength expansion and are related to the conserved angular momentum $L^k$ of the system. This conserved charge can be extracted by the surface integral of the relevant terms in Eqs.~\eqref{EqHGV1} and \eqref{EqHGV2} over the sphere at the spatial infinity. In contrast, the time-dependent part is associated with the GW radiation, whose energy fluxes are of $\mathcal{O}(h^2)$ and decay too fast to reach the far region, which cannot be probed by a distant observer. Hence, we only require Eq.~\eqref{ConV2} be valid up to the leading order in the $(\hat{\mathbf{k}}\cdot \mathbf{y}) \partial_{t_r}$ expansion.  



Now we begin the proof by investigating the following integral of the field $v_i$
\begin{eqnarray}\label{IntVi}
	\hat{k}^j \int_\mathcal{V} d^3y y_j v_i(\tilde{t}_r, \mathbf{y}) = \sum_{n=0}^\infty \frac{1}{n!} \hat{k}^{i_1} \cdots \hat{k}^{i_{n+1}} \int_\mathcal{V}d^3y y_{i_1} \cdots y_{i_{n+1}} \partial^n_{t_r}v_i(t_r, \mathbf{y})\,,
\end{eqnarray}
where the index $j$ is dummy so that we can make the identification $i_{n+1}=j$ on the right-hand side of the equality.  In the following, we shall differentiate the term with $n=0$ and those with $n\geqslant 1$:
\begin{itemize}
	\item $n=0$: The zeroth-order component is given by
	\begin{equation}\label{termV0}
		\hat{k}^j \int_\mathcal{V} d^3y y_j v_i(t_r, \mathbf{y})\,,
	\end{equation}
	which is useful to reproduce the total matter angular momentum later.
	\item $n\geqslant 1$: In each term in this series, there is at least one time derivative acting on $v_i$ so that we can apply the conservation equation in Eq.~\eqref{EqCiV} to these terms
	\begin{eqnarray}\label{Proof1}
		&&\sum_{n=1}^\infty \frac{1}{n!} \hat{k}^{i_1} \cdots \hat{k}^{i_{n+1}} \int_\mathcal{V}d^3y y_{i_1} \cdots y_{i_{n+1}} \partial^n_{t_r}v_i(t_r, \mathbf{y}) \nonumber\\
	   &=& \sum_{n=1}^\infty \frac{1}{n!} \hat{k}^{i_1} \cdots \hat{k}^{i_{n+1}} \int_\mathcal{V}d^3y y_{i_1} \cdots y_{i_{n+1}} \partial^{n-1}_{t_r} \left(\partial_j \partial^j \theta_i -\epsilon_{ijk} \partial^j \omega^k -\partial_{t_r} u_i\right)\,.
	\end{eqnarray}
    We shall analyze terms in the parenthesis. Firstly, we deal with the one involving $\theta_i$:
    \begin{eqnarray}\label{TermTheta}
    	&& \sum_{n=1}^\infty \frac{1}{n!} \hat{k}^{i_1} \cdots \hat{k}^{i_{n+1}} \int_\mathcal{V}d^3y y_{i_1} \cdots y_{i_{n+1}} \partial^{n-1}_{t_r} \partial_j \partial^j \theta_i \nonumber\\
    	&=& \sum_{n=1}^\infty \frac{n+1}{(n-1)!} \hat{k}^{i_1} \cdots \hat{k}^{i_{n-1}} \int_\mathcal{V}d^3y y_{i_1} \cdots y_{i_{n-1}} \partial^{n-1}_{t_r}  \theta_i \nonumber\\
    	&=& 2 \int_\mathcal{V} d^3 y \theta_i (\tilde{t}_r, \mathbf{y}) + \hat{k}^j \int_{\cal V} d^3y y_j \partial_{\tilde{t}_r} \theta(\tilde{t}_r, \mathbf{y})\,,
    \end{eqnarray}
    where the second term in the last equality is exactly the one appearing in Eq.~\eqref{ConV2} but with an opposite sign. Next we consider the term in Eq.~\eqref{Proof1} related to $\omega^k$
    \begin{eqnarray}
    	-\sum_{n=1}^\infty \frac{1}{n!} \hat{k}^{i_1} \cdots \hat{k}^{i_{n+1}} \int_\mathcal{V} d^3y y_{i_1} \cdots y_{i_{n+1}} \partial_{t_r}^{n-1}  \left(\epsilon_{ijk} \partial^j \omega^k \right)\,.
    \end{eqnarray}  
    The first term with $n=1$
    \begin{eqnarray}\label{term1}
    	-\hat{k}^{i_1} \hat{k}^{i_2} \int_\mathcal{V} d^3 y y_{i_1} y_{i_2} \left( \epsilon_{ijk} \partial^j \omega^k \right)
    \end{eqnarray}
    is special, since there is no time derivative acting on $\omega^k$.  We will discuss it later.  For other terms, we can make use of Eq.~\eqref{ConstVe} to represent them in terms of $u_i$ as follows
    \begin{eqnarray}\label{Proof2}
    		&&-\sum_{n=2}^\infty \frac{1}{n!} \hat{k}^{i_1} \cdots \hat{k}^{i_{n+1}} \int_\mathcal{V} d^3y y_{i_1} \cdots y_{i_{n+1}} \partial_{t_r}^{n-1}  \left(\epsilon_{ijk} \partial^j \omega^k \right) \nonumber\\
    		&=& \sum_{n=2}^\infty \frac{1}{n!} \hat{k}^{i_1}\cdots \hat{k}^{i_{n+1}} \int_\mathcal{V} d^3y y_{i_1} \cdots y_{i_{n+1}} \partial^{n-2}_{t_r} \left(\partial^j \partial_j u_i \right) \nonumber\\
    		&=& \sum_{n=2}^\infty \frac{n+1}{(n-1)!} \hat{k}^{i_1} \cdots \hat{k}^{i_{n-1}} \int_{\cal V} d^3y y_{i_1} \cdots y_{i_{n-1}} \partial_{t_r}^{n-2} u_i  \nonumber\\
    		&=& \sum_{n=2}^\infty \frac{1}{(n-2)!} \hat{k}^{i_1} \cdots \hat{k}^{i_{n-1}} \int_\mathcal{V} d^3y y_{i_1} \cdots y_{i_{n-1}} \partial^{n-2}_{t_r} u_i \nonumber\\
    		&& + \sum_{n=2}^\infty \frac{2}{(n-1)!} \hat{k}^{i_1} \cdots \hat{k}^{i_{n-1}} \int_\mathcal{V} d^3y y_{i_1} \cdots y_{i_{n-1}} \partial^{n-2}_{t_r} u_i \,.
    \end{eqnarray}
    The first term in the last equality, when combining with the last term in Eq.~\eqref{Proof1}, gives the following zeroth-order contribution
    \begin{eqnarray}\label{termU0}
    	\hat{k}^j \int_\mathcal{V} d^3 y y_j u_i (t_r, \mathbf{y})\,,
    \end{eqnarray}
    while the second term, we can apply Eq.~\eqref{ConstVe} again to transform it back into the expression of $\omega^k$
    \begin{eqnarray}
    	 &&\sum_{n=2}^\infty \frac{2}{(n-1)!} \hat{k}^{i_1} \cdots \hat{k}^{i_{n-1}} \int_\mathcal{V} d^3y y_{i_1} \cdots y_{i_{n-1}} \partial^{n-2}_{t_r} u_i \nonumber\\
    	 &=& \sum_{n=2}^\infty \frac{2}{(n+1)!} \hat{k}^{i_1} \cdots \hat{k}^{i_{n+1}} \int_\mathcal{V} d^3y y_{i_1} \cdots y_{i_{n+1}} \partial_{t_r}^{n-2} \left(\partial_j \partial^j u_i\right)\nonumber\\
    	 &=& - \sum_{n=2}^\infty \frac{2}{(n+1)!} \hat{k}^{i_1} \cdots \hat{k}^{i_{n+1}} \int_\mathcal{V} d^3y y_{i_1} \cdots y_{i_{n+1}} \partial_{t_r}^{n-1} \left(\epsilon_{ijk} \partial^j \omega^k\right)\,.
    \end{eqnarray}
    Note that the term in Eq.~\eqref{term1} has the same form as those in the summation except for $n=1$, so that we can combine them into a single expression
    \begin{eqnarray}
    	-\sum_{n=1}^\infty \frac{2}{(n+1)!} \hat{k}^{i_1} \cdots \hat{k}^{i_{n+1}} \int_\mathcal{V} d^3y y_{i_1} \cdots y_{i_{n+1}} \partial_{t_r}^{n-1} \left(\epsilon_{ijk} \partial^j \omega^k\right)\,.
    \end{eqnarray}
    By further using the conservation equation of Eq.~\eqref{EqCiV}, we can transform the above expression into the following form
    \begin{eqnarray}\label{TermOmega}
    	&& \sum_{n=1}^\infty \frac{2}{(n+1)!} \hat{k}^{i_1} \cdots \hat{k}^{i_{n+1}} \int_\mathcal{V} d^3y y_{i_1} \cdots y_{i_{n+1}}  \left[\partial^n_{t_r} (v_i+u_i) - \partial^{n-1}_{t_r} \left(\partial^j \partial_j \theta_i\right)  \right]\nonumber\\
    	&=&  \sum_{n=1}^\infty \frac{2}{(n+1)!} \hat{k}^{i_1} \cdots \hat{k}^{i_{n+1}} \int_\mathcal{V} d^3y y_{i_1} \cdots y_{i_{n+1}} \partial^n_{t_r} (v_i+u_i)  - 2 \int_\mathcal{V} d^3y \theta_i(\tilde{t}_r, \mathbf{y})\,.
    \end{eqnarray}
\end{itemize}
By putting together all pieces in Eqs.~\eqref{termV0}, \eqref{TermTheta}, \eqref{termU0} and \eqref{TermOmega}, we can obtain the result of the integral in Eq.~\eqref{IntVi}
\begin{eqnarray}\label{IntVf}
	\hat{k}^j \int_\mathcal{V} d^3 y y_j v_i(\tilde{t}_r, \mathbf{y}) &=& \hat{k}^j \int_\mathcal{V} d^3y y_j (v_i+u_i)(t_r ,\mathbf{y}) + \hat{k}^j \int_\mathcal{V} d^3 y y_j \partial_{\tilde{t}_r} \theta_i (\tilde{t}_r, \mathbf{y})  \nonumber\\
	&& + \sum_{n=1}^\infty \frac{2}{(n+1)!} \hat{k}^{i_1} \cdots \hat{k}^{i_{n+1}} \int_\mathcal{V} d^3y y_{i_1} \cdots y_{i_{n+1}} \partial^n_{t_r} (v_i+u_i) \,,
\end{eqnarray}
where the last term in Eq.~\eqref{TermOmega} cancels the first one in Eq.~\eqref{TermTheta} exactly. Note that there is at least one time derivative acting on $(v_i+u_i)$ in the terms in the second line of  Eq.~\eqref{IntVf}, which means that these terms should be neglected in the long-wavelength limit. 
Consequently, up to the leading order in the long-wavelength expansion, we have
\begin{eqnarray}
	\hat{k}_j \int_{\mathcal{V}} d^3 y y^j (v_i-\partial_t \theta_i) \simeq \hat{k}^j \int_{\cal V} d^3 y y_j (v_i +u_i) (t_r, \mathbf{y}) = -(1/2) \epsilon_{ijk} \hat{k}^j L^k\,,
\end{eqnarray}
which is exactly the consistency relation in Eq.~\eqref{ConV2}. Here we have employed the identity in Eq.~\eqref{idV2} and the definition of the conserved angular momentum $L^k$ in Eq.~\eqref{DefL}.


\bibliography{gwG}

\end{document}